\documentclass[oldversion]{aa}  

\usepackage{graphicx}
\usepackage{txfonts}
\usepackage{natbib}
\bibpunct{(}{)}{;}{a}{}{,}

\newcommand{\ros}{{\em ROSAT}}
\newcommand{\hst}{{\em HST}}

\newcommand{\chan}{{\em Chandra}}

\newcommand{\xmm}{{\em XMM-Newton}}

\newcommand{\eROS}{{\em eROSITA}}
\newcommand{\swi}{{\em Swift}}
\newcommand{\nh}{N_{\rm H}}

\def \msev{M7}

\def \magzersev{RX J0720.4-3125}

\def \magonethr{RX J1308.6+2127}
\def \magtwoone{RX J2143.0+0654}

\def \zersixfiv{2XMM J104608.7-594306}
\def \onesixfou{2XMM J121017.0-464609}
\def \throneeig{2XMM J010642.3+005032}
\def \thrfoueig{2XMM J043553.2-102649}
\def \thrfiveig{2XMM J031459.9-291816}
\def \thrsixfou{2XMM J214026.1-233222}
\def \sixzerfou{2XMM J125904.5-040503}
\def \sixeigone{2XMM J125045.7-233349}

\def \candzsf{XMM J1046}
\def \candosf{XMM J1210}
\def \candtoe{XMM J0106}
\def \candtfoe{XMM J0435}
\def \candtfie{XMM J0314}
\def \candtsf{XMM J2140}
\def \candszf{XMM J1259}
\def \candseo{XMM J1250}

\begin{document}

% ----- HEADER -----------------------------------------------------------------------------------------------
\title{A search for thermally emitting isolated neutron stars\\in the 2XMMp catalogue}
%\title{A search for thermally emitting isolated neutron stars in the 2XMMp catalogue}

\author{A. M. Pires\inst{1,2}
        \and C. Motch\inst{2}
        \and E. Janot-Pacheco\inst{1}
        \thanks{Based on the public data archive of \xmm\, an ESA Science Mission with instruments and contributions directly funded by the ESA Member States and the USA (NASA) and a \chan\ Legacy programme. Optical observations were performed at the European Southern Observatory, Paranal, Chile, under programme ID 079.D-0633(A), and at the Southern Astrophysical Research Telescope, Cerro Pach\'on, Chile.}}

\offprints{A. M. Pires}

\institute{Instituto de Astronomia, Geof\'isica e Ci\^encias Atmosf\'ericas, Universidade de S\~ao Paulo, R. do Mat\~ao 1226, 05508-090 S\~ao Paulo, Brazil,
           \email{apires@astro.iag.usp.br}
           \and
           CNRS, Universit\'e de Strasbourg, Observatoire Astronomique, 11 rue de l'Universit\'e, 67000 Strasbourg, France
}

\date{Received ...; accepted ...}


\titlerunning{A search for INSs in the 2XMMp catalogue}
\authorrunning{A. M. Pires et al.}
% -----------------------------------------------------------------------------------------------------------

% ----- ABSTRACT --------------------------------------------------------------------------------------------
\abstract
{
% Context
The relatively large number of nearby radio-quiet and thermally emitting isolated neutron stars (INSs) discovered in the \ros\ All-Sky Survey, dubbed the ``Magnificent Seven'', suggests that they belong to a formerly neglected major component of the overall INS population. So far, attempts to discover similar INSs beyond the solar vicinity failed to confirm any reliable candidate. 
% Aims
The good positional accuracy and soft X-ray sensitivity of the EPIC cameras onboard the \xmm\ satellite allow to efficiently search for new thermally emitting INSs. We used the 2XMMp catalogue to select sources with no catalogued candidate counterparts and with X-ray spectra similar to those of the Magnificent Seven, but seen at greater distances and thus undergoing higher interstellar absorptions. 
% Method
Identifications in more than 170 astronomical catalogues and visual screening allowed to select fewer than 30 good INS candidates. In order to rule out alternative identifications, we obtained deep ESO-VLT and SOAR optical imaging for the X-ray brightest candidates. We report here on the optical follow-up results of our search and discuss the possible nature of 8 of our candidates. 
% Results
A high X-ray-to-optical flux ratio together with a stable flux and soft X-ray spectrum make the brightest source of our sample, \zersixfiv, a newly discovered thermally emitting INS. The X-ray source \throneeig\ has no evident optical counterpart and should be further investigated. The remaining X-ray sources are most probably identified with cataclysmic variables and active galactic nuclei, as inferred from the colours and flux ratios of their likely optical counterparts. Beyond the finding of new thermally emitting INSs, our study aims at constraining the space density of this Galactic population at great distances and at determining whether their apparently high density is a local anomaly or not.
}
% -----------------------------------------------------------------------------------------------------------

\maketitle

% ----- INTRODUCTION ----------------------------------------------------------------------------------------
\section{Introduction}

Around ten years ago, seven X-ray bright thermally emitting and radio-quiet isolated neutron stars (INSs) sharing similar properties were identified in the \ros\ All-Sky Survey data. This group is now commonly referred to as the ``Magnificent Seven'' (or \msev, for simplicity; see \citealt[][for a review]{hab07}) for the reasons that they clearly stand apart from the population of rotation-powered radio pulsars. In spite of many searches for similar objects in the \ros\ data \citep[e.g.][]{rut03,chi05,agu06}, no new candidate was identified since the discovery of the last member, \magtwoone\ \citep{zam01}.

A soft blackbody spectrum undergoing low interstellar absorption ($kT\sim40$\,--\,100\,eV and $\nh\sim$ few $10^{20}$\,cm$^{-2}$) is common to the seven sources, as well as the absence of a non-thermal component extending towards higher energies. We note, however, the detection in the optical/UV of a possible non-thermal power-law component in the spectrum of \magzersev\ \citep{kap03}. Very faint optical counterparts with blue magnitudes of $m_B\sim25$\,--\,28 were detected for several sources, implying high logarithmic X-ray-to-optical flux ratios of $\sim$\,4\,--\,5. Relatively to radio pulsars, the neutron star spin periods are longer and distributed in a much narrower range, $P\sim3$\,--\,10\,s. Six of the sources show sinusoidal X-ray pulsations with pulsed fractions between $\sim$\,1\% and 18\%.
The \msev\ are believed to be nearby ($d\la500$\,pc), as inferred from the distribution of the interstellar medium in the line-of-sight and the equivalent hydrogen column densities measured in their X-ray spectra \citep{pos07}. Furthermore, \hst\ parallaxes are available for the two X-ray brightest members \citep{kap02,kap07} with results largely consistent with those estimated by \cite{pos07}. The seven sources do not show either persistent or transient radio emission to a rather sensitive limiting flux of $\sim10$\,$\mu$Jy \citep{kon08} and are not associated to supernova remnants. However, unconfirmed claims of detection at long wavelengths exist \citep{mal07}.

Proper motion measurements in the optical (for the three INSs with bright enough counterparts, \citealt{neu01,kap02,mot03,kap07,mot05,zan06}) and in X-rays (for \magonethr, \citealt{mot08,mot09}) have established these objects as cooling, middle-aged ($\sim$\,10$^5$\,--\,10$^6$\,yr) neutron stars, probably originated in the nearby OB associations of the Gould Belt. 
Intense magnetic fields $B\sim10^{13}$\,--\,10$^{14}$\,G \citep[see][for reviews]{hab07,ker07a} are inferred from the presence of one or more broad absorption lines in the spectra of the sources and from the spin-down rates detected in three cases by timing studies \citep{kap05a,kap05b,ker08}, where it is assumed magnetic dipole braking. For these cases, characteristic ages $\tau_{\rm{ch}} = P/2\dot{P}\sim$ 1.5\,--\,4\,Myr overpredict by a factor of $\sim$\,2\,--\,10 the kinematic travel times inferred by tracing back the trajectories from their likely birth places \citep[e.g.][]{mot08}. The spectral absorption lines are commonly interpreted as due to atomic transitions in a partially ionised hydrogen atmosphere \citep[e.g.][]{lai01} or as proton cyclotron features \citep[e.g.][]{zan01}.
In spite of the high $B$, their low magnetospheric activity allows for a direct view of the surface of the neutron star. Atmosphere models and the surface emissivity under extreme physical conditions can thus be tested in order to derive radii \citep[e.g.][]{hok07} and eventually constrain the equation of state of neutron star interior, provided that distances are known. Unfortunately, the current lack of understanding of the surface composition, magnetic field and temperature distributions have limited any definite conclusion so far.

The locus occupied by the \msev\ in the $P$\,--\,$\dot{P}$ diagram, somewhat intermediate between radio pulsars and magnetars, suggests that they could be linked to other populations of INSs. In particular, the similar spin periods and intense magnetic fields raised the possibility that some of the \msev\ could have evolved from the younger and more energetic magnetar objects \citep{hey98}. Magnetic field decay provides an additional source of heating of the neutron star crust and dramatically changes the cooling history of neutron stars with $B\ga10^{13}$\,G. \cite{agu08} have shown that the range of observed temperatures of the \msev\ is consistent with them being born as magnetars with $B\sim10^{14}$\,--\,$10^{15}$\,G, provided that their current ages are around $10^6$\,yr (which is in agreement with the kinematic ages). 
On the other hand, if the neutron star magnetic field has not changed substantially over its lifetime, an evolutionary link with the high $B$ radio pulsars (HBPSRs) can be suggested \citep[e.g.][]{zan02}. Relatively to the \msev, these objects are generally located at much greater distances and show considerably higher spin-down energy ($\dot{E}$). 
However, if the HBPSRs are a factor of $\sim$\,100 younger, then the difference in $\dot{E}$ can be explained within the standard scenario for pulsar evolution assuming the usual magnetic dipole braking model \citep{kap08}. 
The \msev\ could then be long-period radio pulsars for which the narrow emission beam simply does not sweep over the earth.
Finally, the only rotating radio transient (RRAT) discovered so far at X-ray energies \citep{lau07} also shares strikingly similar properties with the \msev\ -- namely the position on the $P$\,--\,$\dot{P}$ diagram and the presence of a broad absorption line in its thermal spectrum. The source is however hotter ($kT\sim140$\,eV), fainter and expected to be more remote, at $\sim$\,3.6\,kpc.

Considering that within 1\,kpc the \msev\ appear in comparable numbers as young ($\la$ few Myr) radio and $\gamma$-ray pulsars \citep{pop03}, they may represent the only identified members of a large, yet undetected, elusive population of radio-quiet and thermally emitting INSs in the Galaxy. It should be noted, however, that the proximity of OB associations of the Gould Belt to the solar vicinity is an important factor that might explain this apparent local overdensity of INSs with similar temperatures, ages and magnetic field intensities. Since the neutron star cooling is strongly dependent on mass and, to a lower extent, on magnetic field, a scenario where these stars evolved from a common progenitor population of massive stars could be considered.
The discovery of similar sources at greater distancies is then mandatory in order to make any progress towards understanding their population properties and evolutionary links with other groups of Galactic INSs.

Despite the relatively small field-of-view and low sky coverage of the \xmm\ Observatory, its large effective area and good positional accuracy at soft X-ray energies make it ideal to look for faint INS candidates beyond the Gould Belt. In \cite{pir08a} we reported on the preliminary results of a programme aimed at identifying new thermally emitting INSs in the 2XMMp catalogue. In the present paper we discuss the nature of a handful of the X-ray brightest of our INS candidates, selected from more than 7.2\,$\times$\,10$^4$ EPIC pn sources with count rates above 0.01\,s$^{-1}$.
The paper is structured as follows: in Section~\ref{sec_selec} we describe the methodology which was applied to select the 2XMMp sources and restrict the number of INS candidates; in Section~\ref{sec_fllup} the optical data obtained on a subsample of our candidates are described together with their analysis and results; Section~\ref{sec_X} is devoted to the analysis of the X-ray emission of these candidates. Finally, the discussion, conclusions and summary are given in Sections~\ref{sec_disc} and \ref{sec_conc}. The discovery of the new thermally emitting INS \zersixfiv, the X-ray brightest among our sample, and the discussion of its nature as, in particular, a cooling or an accreting neutron star have been presented in \cite{pir08b}.
% -----------------------------------------------------------------------------------------------------------

% ----- SECTION - SELECTION ---------------------------------------------------------------------------------
\section{Selection of candidates\label{sec_selec}}

% ----- TABLE - SELECTION 2XMMp -----------------------------------------------------------------------------
\begin{table}[t]
\begin{center}
\caption{Results of visual screening of the selected 2XMMp sources discriminated by selection region\label{tab_2xmmpsel}}
\begin{tabular}{l c c c c} 
\hline \hline
 & I & II & III & IV\\
\hline
Known INSs                                & 16 &  0 &  1 &  1 \\
INS candidates                            &  1 &  5 & 24 &  2 \\
Solar system objects                      &  2 &  0 &  0 &  0 \\
Extragalactic X-ray sources               & 12 & 31 & 39 &  6 \\
Bright knots in SNRs                      &  9 & 59 & 18 & 51 \\
Diffuse emission in clusters of galaxies  &  2 &  6 &  9 &  0 \\
Spurious out-of-time events / edge        & 10 & 70 & 58 &  8 \\
Spurious in wing of bright object         &  2 & 15 & 18 &  6 \\
Sources in CCD gap                        &  3 & 13 & 46 &  5 \\
Other                                     &  2 &  3 &  8 &  0 \\
\hline
Total                                     & 59 & 202& 221& 79 \\
\hline
\end{tabular}
\end{center}
\end{table}
% -----------------------------------------------------------------------------------------------------------

% ----- FIGURE - HR DIAGRAM ---------------------------------------------------------------------------------
\begin{figure*}
\centering
\includegraphics[width=0.495\textwidth]{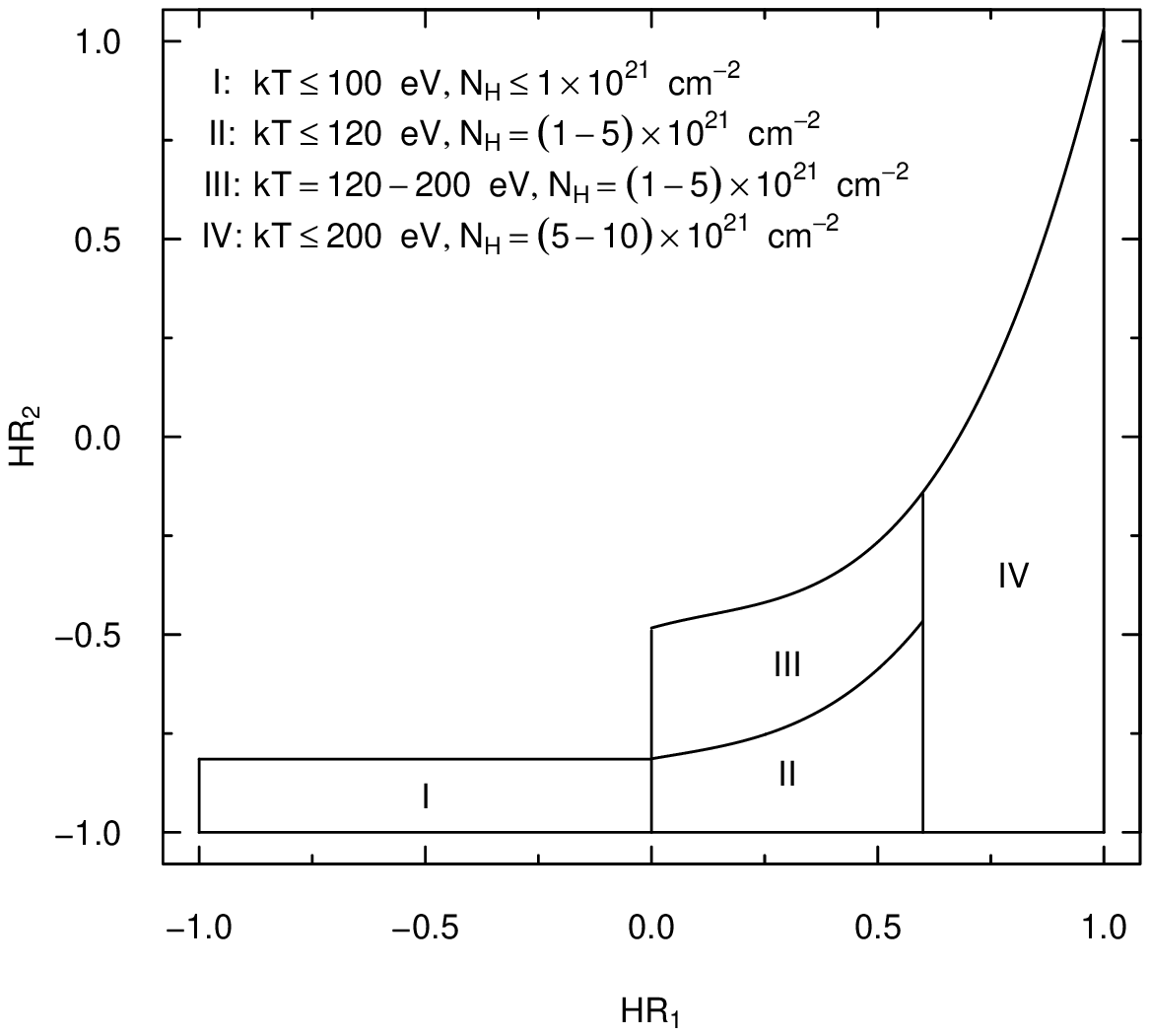}
\includegraphics[width=0.495\textwidth]{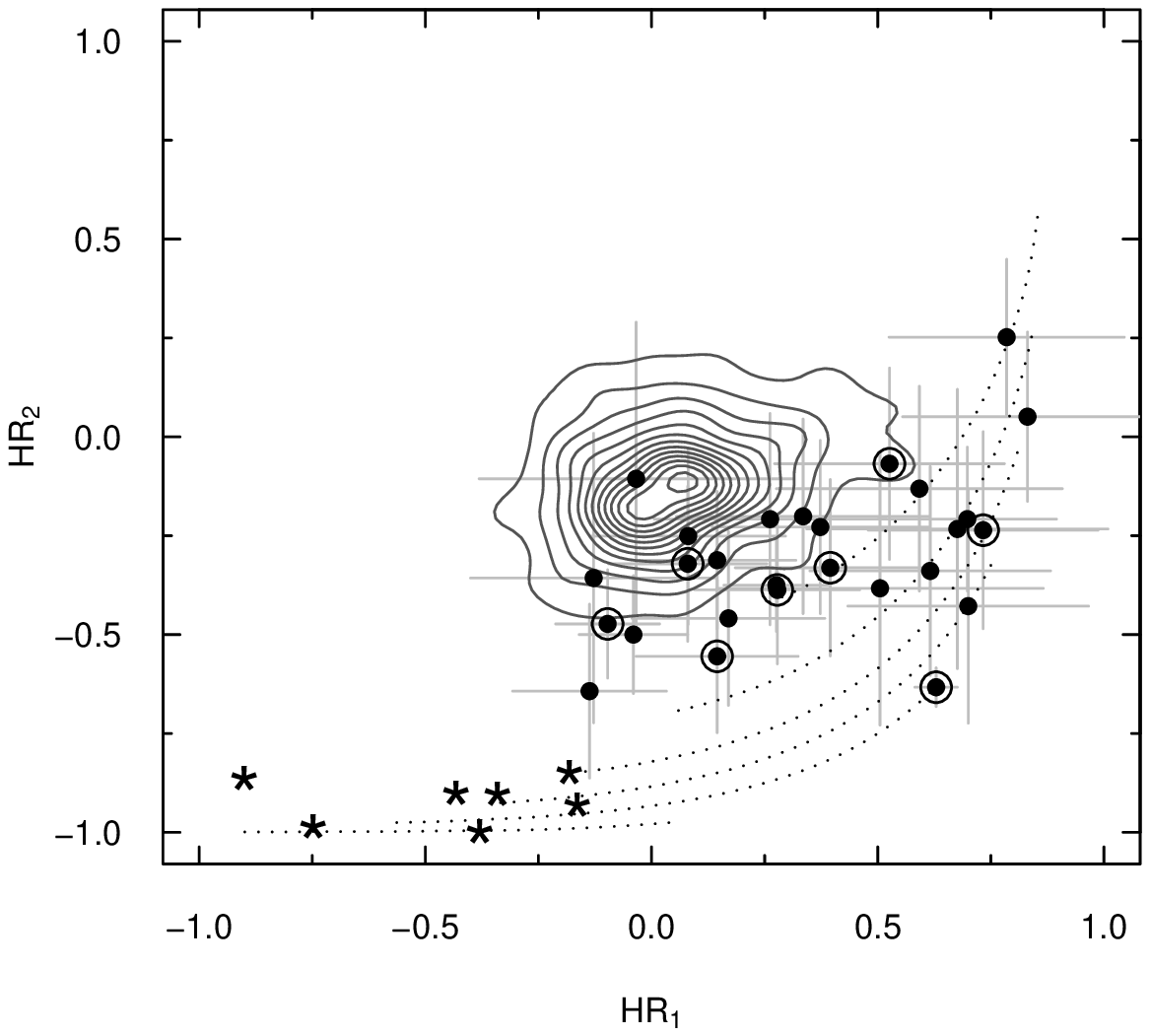}
\caption{\textit{Left:} Boundaries of the 4 different selection regions considered in the HR$_1$\,$\times$\,HR$_2$ diagram. \textit{Right:} Positions of the final list of INS candidates in the same diagram (filled circles with error bars). The known \ros-discovered INSs (stars) occupy the less absorbed and cooler portion of the diagram. Dotted lines denote soft absorbed blackbodies of different temperatures (50\,eV, 80\,eV, 100\,eV, 120\,eV, 150\,eV and 200\,eV, from bottom to top) undergoing hydrogen column densities in range 10$^{19}$ to 10$^{22}$\,cm$^{-2}$. Contours show the hardness ratio distribution of quasars from SDSS DR3 (see text). The candidates selected for follow-up optical observations are highlighted by open circles.\label{fig_bestcand}}
\end{figure*}
% -----------------------------------------------------------------------------------------------------------
We searched for new thermally emitting INSs similar to the \msev\ in the pre-release of the \xmm\ serendipitous source catalogue, 2XMMp\footnote{\texttt{http://xmmssc-www.star.le.ac.uk/Catalogue/\\xcat\_public\_2XMMp.html}}. This version of the catalogue, released in July 2006, contains source detections drawn from 2400 individual \xmm\ EPIC observations made between 2000 and 2006. It contains more than 120,000 unique X-ray sources with a median flux of 2.4\,$\times$\,10$^{-14}$\,erg\,s$^{-1}$\,cm$^{-2}$ (in the total energy band 0.2\,--\,12\,keV); around 20\% of the catalogued sources have fluxes below 1\,$\times$\,10$^{-14}$\,erg\,s$^{-1}$\,cm$^{-2}$. Detection positional accuracy is generally $<$\,$2''$ (68\% confidence radius) and the total sky area covered is $\sim$\,285\,deg$^{2}$.
Despite the small percentage of sky coverage (0.7\%), the EPIC instruments provide much better positional accuracy at soft energies and faint fluxes relative to \ros, making the \xmm\ catalogue a powerful tool to look for unidentified INSs. In the near future, the \ros\ successor \eROS\footnote{\texttt{http://www.mpe.mpg.de/projects.html\#erosita}} mission should perform the first imaging all-sky survey up to 10\,keV with unprecedented spectral and angular resolution, and can potentially increase the number of cooling INSs up to one order of magnitude \citep{pos08}.

In order to select candidates, we only considered detections in the EPIC pn camera at a limiting count rate of 0.01\,s$^{-1}$ (0.2\,--\,12\,keV), corresponding to a total number of $\sim$\,7.2\,$\times$\,10$^4$ X-ray sources, and then required these to be well detected and point-like. Translating into the more precise catalogue parameters, we selected sources with a maximum likelihood of detection PN\_DET\_ML $>8$ and required a low probability for extension (EP\_EXTENT\_ML $<4$), both properties computed by the catalogue pipeline with the SAS\footnote{\texttt{http://xmm.esac.esa.int/sas}} task \textsf{emldetect}. None of the source quality or variability flags provided by the catalogue was used in the selection and screening process. 

For the remaining $\sim$\,4.6\,$\times$\,10$^4$ entries, we then required as main selection criteria ($i$) no correlations (within 3\,$\sigma$) with the USNO-A2.0 optical catalogue and ($ii$) a soft energy distribution. Applying the first criterion led to the rejection of more than 2.1\,$\times$\,10$^4$ X-ray sources having a positive match with USNO-A2 entries. For requirement ($ii$), instrumental response and auxiliary files were used to simulate the count rates and hardness ratios (HR) of absorbed blackbodies (in a grid of temperature and column density), taking into account the different optical blocking filters. Although the effective area of the three (thin, medium and thick) filters at energies below 1\,keV can differ by a factor of up to 3, we verified that their influence in the HR diagrams is minor. 

The HR is defined as
\begin{equation}
\textrm{HR}_i = \frac{C_{i+1} - C_i}{C_{i+1} + C_i}\qquad i=1,\dots,4
\end{equation}
where $C_i$ and $C_{i+1}$ are source counts into two contiguous of the five pre-defined energy bands of the 2XMMp catalogue, 0.2\,--\,0.5\,keV, 0.5\,--\,1\,keV, 1\,--\,2\,keV, 2\,--\,4.5\,keV and 4.5\,--\,12\,keV. 
We thus selected the catalogue sources with X-ray energy distribution compatible (within 1\,$\sigma$) with that of template blackbodies of $kT\le200$\,eV undergoing absorptions in range $\nh = 10^{19}-10^{22}$\,cm$^{-2}$, by defining their positions in HR diagrams.

The X-ray column densities of the catalogue sources were then verified to be less than or equal to the total Galactic value in order to discard intrinsically absorbed objects. Sources with very large positional uncertainties ($r_{\rm 90}>4''$, where $r_{\rm 90}$ is the source 90\% confidence level error circle) or which were too far from the optical axis (at off-axis angles $\theta > 11'$) were also discarded. After applying these selections, we ended up with $\sim$\,1000 sources. Of these, a total additional number of 461 X-ray sources with matches in one or more of the optical/IR catalogues USNO-B1.0, SDSS, GSC-2, APM and 2MASS was discarded as well. We note that our main sources of information on the possible astronomical content of the X-ray source error circles were the cross-correlations computed by the Astronomical Catalogue Data Subsystem. This module of the \xmm\ reduction pipeline \citep{wat09} provides lists of archival entries in over 170 archival catalogues having a position consistent within 3\,$\sigma$ with that of the 2XMMp source.

So as to more easily classify the 561 remaining sources which fulfilled the above selection criteria, we considered 4 different regions in the HR$_1$\,$\times$\,HR$_2$ diagram, based on temperature and column density (see left panel of Fig.~\ref{fig_bestcand}). We note that, although this diagram represents our main means of selecting INS candidates, we required soft emission in the HR$_3$\,$\times$\,HR$_4$ diagram as well so as to avoid X-ray sources with significant emission above 2\,keV. For every source in these regions, we visually checked the X-ray and optical images\footnote{Finding charts from SDSS or created from digitized optical plates.} and searched for possible identifications in astronomical catalogues and databases like NED and Simbad. Source selection and screening processes used the facilities provided by the XCat-DB \citep{mot07b}, a database hosting the 2XMM catalogue and its associated pipeline products, including the archival cross-correlations.

Many of the sources turned out to be false detections in extended diffuse emission of mostly supernova remnants (SNRs) or due to out-of-time events. The high number of spurious detections among our selected sources reflects the fact that we did not use the source quality flag information in the early steps of the selection process. X-ray sources located in the direction of nearby galaxies (sometimes with an optical or infrared counterpart visible in the finding charts) were classified as ``extragalactic'' and thus were no longer regarded as potential INS candidates. Finally, because the spectral properties of some faint X-ray sources are not well determined in short exposures, the last step in our procedure was to check if any source with no evident classification (and thus a potential INS candidate) was spectrally harder in other (longer) \xmm\ exposures. 

% ----- FIGURE - HISTOGRAMS ---------------------------------------------------------------------------------
\begin{figure}
\centering
\includegraphics[width=0.475\textwidth]{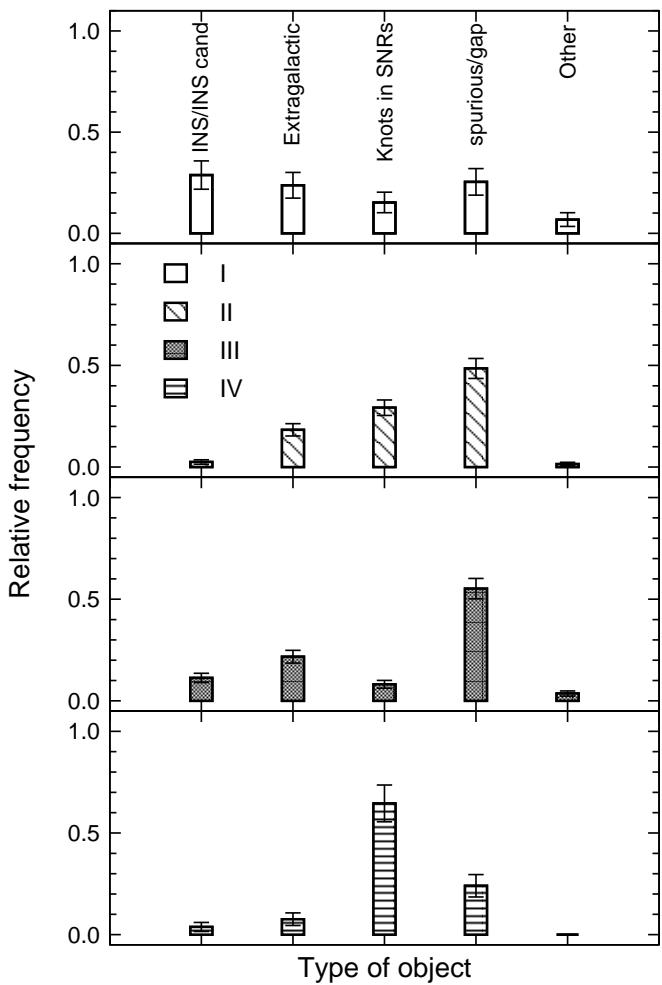}
\caption{Histograms showing the relative frequency of the 2XMMp sources that fulfilled the selection criteria applied in order to find new INS candidates. The four plots show the same histograms in the different selection regions (see Fig.~\ref{fig_bestcand}).\label{fig_hist}}
\end{figure}
% -----------------------------------------------------------------------------------------------------------
In Table~\ref{tab_2xmmpsel} we list the results of the final screening of sources, sorted by selection regions. The numbers refer to detections that might have entered more than once one or more selection regions and not to unique sources\footnote{In other words, a soft source which was observed several times by the \xmm, like one of the \msev, can enter the same selection region more than once, provided that its spectral properties are compatible. Similarly, one INS candidate can show HR that are consistent with more than one selection region, and in this case it is counted accordingly.}. It is worth noting that the known \msev\ or, more precisely, the pn observations of the \msev\ included in the compilation of the 2XMMp were among the final list of INS candidates, providing a test of the efficiency of our procedure in selecting thermally emitting neutron stars.

Overall, we found 59 very soft and low-absorbed sources with X-ray emission consistent with blackbodies of temperatures $kT\le100$\,eV and column densities $\nh\le1$\,$\times$\,10$^{21}$\,cm$^{-2}$ (region I in Fig.~\ref{fig_bestcand}). In this region we retrieved the totality (16) of pn observations of the \msev\ carried out until 2006. Two observations of Saturn and a number of bright knots in the extended diffuse emission of Galactic SNRs (of the supernova SN 1006 and of the relatively old and nearby Cygnus Loop) also are included in this region. Interestingly, we noted that there were some cases (included as ``other'' in Table~\ref{tab_2xmmpsel}) in which the X-ray source had no optical candidate but was nonetheless related to (optically bright) high proper motion (HPM) stars. For these cases, we checked that the position of the X-ray source was compatible with the position of the HPM star at the time of the X-ray observation. Region I provided 1 INS candidate. 

More absorbed and slightly hotter sources, consistent with blackbodies of $kT\le120$\,eV and $\nh=(1$\,--\,$5)\times10^{21}$\,cm$^{-2}$ (region II), totalized a number of 202. This selection region provided 4 INS candidates (in addition to the INS candidate that had also been selected in region I).
Most of the sources consisted of bright knots in SNRs -- the Galactic remnants of Puppis, RCW 86 and SN 1006 as well as SNRs in galaxies M33, M31, M51 and the Small Magellanic Cloud -- and of spurious detections (mostly of bright objects on the edge of the CCDs or sources located in the CCD gaps) and of out-of-time events.

% ----- TABLE - PROPERTIES CANDIDATES -----------------------------------------------------------------------
\begin{table*}
\begin{minipage}[t]{\textwidth}
\caption{Properties of the INS candidates selected for optical follow-up\label{tab_cand}}
\centering
\renewcommand{\footnoterule}{}
\begin{tabular}{l l l c c c r c c} 
\hline \hline
 Source identification & RA & DEC & $b$\footnote{Galactic latitude.} & $\nh^{\rm Gal}$\footnote{Total Galactic extinction \citep{dic90}.} & $r_{90}$\footnote{Positional error (90\% confidence level).} & pn CR\footnote{EPIC pn count rate in total energy band (0.2\,--\,12\,keV).} & Selection & Optical\\
 & (J2000) & (J2000) & (degrees) & (cm$^{-2}$) & (arcsec) & (s$^{-1}$) & Region & Data\footnote{Obtained during different observing periods: $^\star$ESO-VLT P79 (pre-imaging), $^\dag$SOAR 2007A, $^\ddag$SOAR 2007B, $^\diamondsuit$SOAR 2008A.}\\
\hline
\zersixfiv & 10 46 08.7 &  -59 43 06.1 & -0.60  & 1.35\,$\times$\,10$^{22}$ & 1.33 &  0.060(4)   & II, III, IV & $\star,\diamondsuit$\\ 
\onesixfou & 12 10 17.1 &  -46 46 11.2 & +15.52 & 8.70\,$\times$\,10$^{20}$ & 2.60 &  0.027(6)   & III         & $\star$\\ 
\throneeig & 01 06 42.4 &  +00 50 31.3 & -61.79 & 3.18\,$\times$\,10$^{20}$ & 3.80 &  0.020(5)   & III         & $\star$\\ 
\thrfoueig & 04 35 53.2 &  -10 26 50.0 & -34.80 & 5.80\,$\times$\,10$^{20}$ & 2.70 &  0.019(3)   & III         & $\ddag$\\ 
\thrfiveig & 03 14 59.9 &  -29 18 15.5 & -58.42 & 1.32\,$\times$\,10$^{20}$ & 2.41 &  0.018(4)   & III         & $\ddag$\\ 
\thrsixfou & 21 40 26.2 &  -23 32 22.3 & -46.95 & 3.51\,$\times$\,10$^{20}$ & 1.90 &  0.0181(20) & III         & $\dag$\\ 
\sixzerfou & 12 59 04.6 &  -04 05 02.3 & +58.73 & 1.81\,$\times$\,10$^{20}$ & 1.90 &  0.0129(21) & III         & $\dag$\\ 
\sixeigone & 12 50 45.7 &  -23 33 47.7 & +39.31 & 7.37\,$\times$\,10$^{20}$ & 3.30 &  0.0122(21) & III         & $\dag$\\ 
\hline
\end{tabular}
\end{minipage}
\end{table*}
% -----------------------------------------------------------------------------------------------------------

A total of 221 sources were found in region III ($kT=120$\,--\,200\,eV and $\nh=(1$\,--\,$5)\times10^{21}$\,cm$^{-2}$). Relatively to region II, a less important fraction of the total number of selected sources consisted of spurious detections in SNRs, which in this region were mostly extragalactic. Also relatively to region I and II, a larger number of extragalactic X-ray sources were found. Region III also provided the largest number of INS candidates (22, in addition to 2 repeated entries of selection regions I and II). Interestingly, we detected in this region the X-ray emission of pulsar PSR J1722-3712 and the quiescent neutron stars XMMU J164143.8+362758 and CXOU J132619.7-472910 in the globular clusters M13 and $\omega$ Cen. These two sources were included as ``other'' in Table~\ref{tab_2xmmpsel} while the pulsar was included as ``known INSs''.

Finally, the most absorbed sources, with X-ray emission consistent with $kT\le200$\,eV and $\nh=(5$\,--\,$10)\times10^{21}$\,cm$^{-2}$ (region IV), corresponded to a number of 79. The detections mainly consisted of knots in the Galactic SNRs Puppis, RCW 86, RCW 89 and Tycho as well as in some extragalactic SNRs in the Small and Large Magellanic Clouds. This region provided no new INS candidate -- the objects in Table~\ref{tab_2xmmpsel} corresponding to an INS candidate which already had been selected in region II and to the pulsar found in region III. The results sorted by selection region can be visualized in the histograms of Fig.~\ref{fig_hist}.

In short, out of 7.2\,$\times$\,10$^4$ serendipitous EPIC pn sources above 0.01\,s$^{-1}$, fewer than 30 candidates met all the selection criteria; these are thus intrinsically soft sources not associated to any catalogued optical or infrared object and not likely to be spurious. In the right panel of Fig.~\ref{fig_bestcand} are shown the positions of the final list of unique INS candidates in the hardness ratio diagram HR$_1$\,$\times$\,HR$_2$. The lowest left part of the diagram is occupied by the soft, low-absorbed \msev\ while our candidates, undergoing higher photoelectric absorptions, move upwards, along with the blackbody lines of hotter temperatures. Contour lines in the same diagram show the hardness ratio distribution of quasars from the SDSS DR3 \citep{schneider05} which have counterparts in the 2XMMi catalogue with a maximum likelihood of detection greater than 20 (total of 796 sources; the list of correlated sources was extracted using the XCat-DB). It is worth noting that their location is right above our selection region III. We can see already that several of our INS candidates have hardness ratios compatible with those of the SDSS quasars.

% ----- TABLE - OPTICAL OBSERVATIONS ------------------------------------------------------------------------
\begin{table*}
\caption{Description of the optical observations\label{tab_logopt}}
\begin{center}
\begin{tabular}{l c c l l l l l} 
\hline \hline
Target & Night & Telescope & Exposures & Exp. Time & Total Exp. Time & FWHM & Airmass\\
 & & & & (s) & (s) & (arcsec) & \\
\hline
\candzsf  & 22-02-2007 & ESO-VLT  & 2\,$\times$\,B, 2\,$\times$\,R                  & 150, 150      & 300, 300        & 0.81, 0.68       & 1.27, 1.26\\
\candosf  & 25-02-2007 & ESO-VLT  & 2\,$\times$\,B, 2\,$\times$\,R                  & 150, 150      & 300, 300        & 0.83, 0.70       & 1.16, 1.15\\
\candseo  & 09-05-2007 & SOAR & 14\,$\times$\,U, 7\,$\times$\,B, 3\,$\times$\,R & 500, 200, 200 & 7000, 1400, 600 & 0.90, 0.74, 0.80 & 1.10, 1.03, 1.01\\
\candszf  & 10-07-2007 & SOAR & 4\,$\times$\,U, 4\,$\times$\,B, 4\,$\times$\,R  & 500, 200, 200 & 2000, 800, 800  & 0.75, 0.68, 0.62 & 1.02, 1.01, 1.01\\
\candtsf  & 11-07-2007 & SOAR & 4\,$\times$\,U, 4\,$\times$\,B, 3\,$\times$\,R  & 500, 200, 200 & 2000, 800, 600  & 1.11, 0.95, 0.80 & 1.19, 1.23, 1.25\\
\candtoe  & 21-07-2007 & ESO-VLT  & 2\,$\times$\,B, 2\,$\times$\,R                  & 150, 150      & 300, 300        & 0.62, 0.58       & 1.16, 1.15\\
\candtfie & 29-01-2008 & SOAR & 4\,$\times$\,U, 5\,$\times$\,B, 4\,$\times$\,R  & 500, 200, 200 & 2000, 1000, 800 & 0.85, 0.82, 0.72 & 1.18, 1.22, 1.24\\
\candtfoe & 30-01-2008 & SOAR & 4\,$\times$\,U, 4\,$\times$\,B, 4\,$\times$\,R  & 500, 200, 200 & 2000, 800, 800  & 1.14, 1.16, 1.10 & 1.10, 1.11, 1.12\\
\candzsf  & 08-02-2008 & SOAR & 6\,$\times$\,V                                  & 1120 & 6720 & 0.94 & 1.23\\
\hline
\end{tabular}
\end{center}
\end{table*}
% -----------------------------------------------------------------------------------------------------------
We list in Table~\ref{tab_cand} the equatorial coordinates and count rates of a subsample of the candidates, which were object of follow-up observations in the optical during the years of 2007 and 2008. The choice of sources for the follow-up was made by selecting the X-ray brightest INS candidates which were visible from the southern hemisphere. Positions are derived by the catalogue pipeline with the SAS task \textsf{emldetect}. The 90\% confidence level error circle on the position is given by $r_{90} = 2.15\sqrt{\sigma^2 + \sigma_{\rm{syst}}^2}$, where $\sigma$ is the nominal error as given by \textsf{emldetect}. The systematic error $\sigma_{\rm{syst}}$ on the detection position is provided by the 2XMMp catalogue and varies with the success or failure of boresight correction of the EPIC sources with optical matches in the USNO-B1.0 catalogue. 
% -----------------------------------------------------------------------------------------------------------

% ----- SECTION - FOLLOW UP ---------------------------------------------------------------------------------
\section{Optical follow-up\label{sec_fllup}}

We are presently conducting an optical campaign with the purpose of investigating the nature of the X-ray brightest INS candidates of our sample. The immediate goal is to find a possible alternative identification -- essentially faint polar-type cataclysmic variables (CVs), late-type stars or active galactic nuclei (AGN) -- to the selected sources, through the analysis of the spectra and the colour indexes of the optical objects that may be present in the X-ray error circles.

The 8 INS candidates in Table~\ref{tab_cand} were observed using the 8.2\,m ESO Very Large Telescope (ESO-VLT) and the 4.1\,m Southern Astrophysical Research Telescope (SOAR) facilities in Chile during 2007 and 2008. Table~\ref{tab_logopt} shows the log of all the observations that we have obtained and analysed so far. We note that the exposure times, seeing and airmasses listed in Table~\ref{tab_logopt} are averaged per filter.
Hereafter, we adopt the convention ``XMM J\textit{hhmm}'' to designate the 8 X-ray sources discussed in this work. 

\subsection{Observations and data reduction}
We obtained deep imaging of the fields of our candidates under photometric sky conditions (see the log of observations in Table~\ref{tab_logopt}). The FOcal Reducer low/dispersion Spectrograph (FORS2; \citealt{app98}) and SOAR Optical Imager (SOI; \citealt{schwarz04}) were used. The FORS2 detector consists of a mosaic of two 2\,k\,$\times$\,4\,k MIT CCDs (15\,$\mu$m) and is optimized for the red band. It provides imaging at a pixel scale of $0.25''$\,pixel$^{-1}$ using standard read-out mode and a 2\,$\times$\,2 binning ($6.8'\times6.8'$ field-of-view). With this configuration the gain is 1.25\,e$^{\textrm{-}}$\,ADU$^{-1}$ and the read-out noise 2.7\,e$^{\textrm{-}}$. The SOI instrument uses a mosaic of two E2V 2\,k\,$\times$\,4\,k CCDs (15\,$\mu$m) and it is optimized for the blue and UV bands. It has a field-of-view of $5.3'\times5.3'$ and a pixel scale of $0.15''$\,pixel$^{-1}$ using a 2\,$\times$\,2 binning. It was used in slow read-out mode, minimizing the read-out noise to 3.1\,e$^{\textrm{-}}$ with a gain of 0.4\,e$^{\textrm{-}}$\,ADU$^{-1}$.

Only pre-imaging data\footnote{The pre-imaging had the purpose of selecting potential optical candidates for spectroscopy or deep imaging.} were obtained with the ESO-VLT while $\sim$\,90\% of the proposed observing time was executed by SOAR in three different observing periods. Dithering patterns with offsets of $2''$ and $5''$ were chosen for the SOAR and the ESO-VLT observations, respectively. Seeing during the several nights of observation varied between $0.58''$ and $1.2''$\,FWHM, with mean values of $0.70''$ and $0.88''$ for the ESO-VLT and SOAR nights.

We used the already bias and flat field corrected frames provided by the ESO-VLT pipeline in the analysis. Images in each filter were combined using an algorithm to remove cosmic rays and bad pixels based on CCD statistics (\textsf{ccdclip} in IRAF\footnote{\texttt{http://iraf.noao.edu}} task \textsf{imcombine}). SOAR data were reduced using standard procedures within IRAF V2.14 \citep{tod86}. The observations were bias and overscan subtracted and combined skyflat and dome flat images in each filter were used to correct the pixel-to-pixel response variations across the CCDs. The science frames in each filter were also combined to eliminate cosmic-rays and increase the signal-to-noise ratio of the data.

The observation of photometric standard stars fields \citep{lan92} was arranged to be executed during the same night as our SOAR targets. Fields RU 149, Mark A and PG1323-085 were selected to calibrate the observations of sources \candtfoe, \candtfie, \candtsf, \candszf\ and \candseo. The deeper observations of \candzsf\ were calibrated with the standards PG1047+003 and SA 104. For every field (as appropriate depending on the requested observations of the science target) we obtained 10\,s (U or V) and 5\,s (B or R) single exposures. The airmasses at which the observations of standard stars were carried out were similar to the ones of the science targets. As only one exposure was taken in each filter (at a given airmass) it was not possible to determine the extinction coefficients; mean values for the periods were adopted instead. On the other hand, the zero-point magnitudes and colour coefficients were determined from our observations of the standard stars in each night. The fit to the standard system was performed relative to the $U-B$ (for the $U$ magnitudes) and $B-R$ (for the $B$ and $R$ magnitudes) colours. The $V$ SOAR instrumental magnitudes (period 2008A) were transformed adopting a null colour coefficient and the mean value of the extinction coefficient for the period. The instrumental magnitudes of the ESO-VLT images were transformed using directly the extinction coefficients and colour terms provided by the ESO-VLT calibration web pages\footnote{\texttt{http://www.eso.org/observing/dfo/quality}}. We compared the R magnitudes of non-saturated GSC-2 stars present in the fields with our results and concluded that the agreement is good with a mean dispersion of $\sim$\,0.03\,mag.

\subsection{Data analysis and results}

\subsubsection{Astrometry}
The astrometric calibration was performed using the USNO-B1.0, 2MASS and GSC-2 catalogues and the GAIA 4.2-1 software\footnote{\texttt{http://star-www.dur.ac.uk/$\sim$pdraper/gaia/gaia.html}}. The pixel coordinates of the non-saturated catalogued stars in every combined science frame were determined by fitting a two-dimensional Gaussian function to their intensity profiles. An astrometrical solution was then computed by fitting the pixel to the celestial coordinates. Overall, our astrometric errors are of $\sim$\,$0.15''$ or better.

% ----- FIGURE - FINDING CHARTS -----------------------------------------------------------------------------
\begin{figure*}
\centering
\includegraphics[width=0.2425\textwidth]{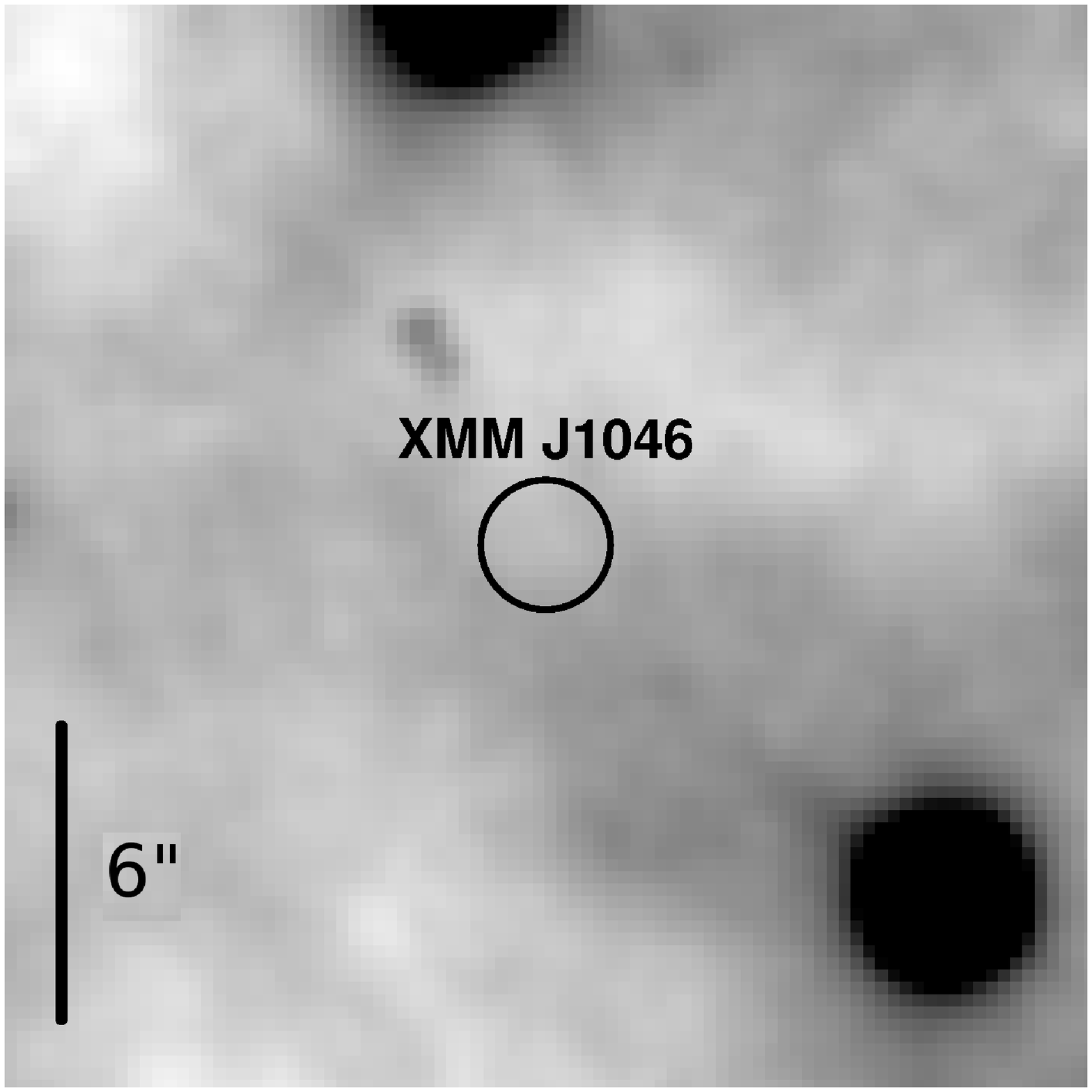}
\includegraphics[width=0.2425\textwidth]{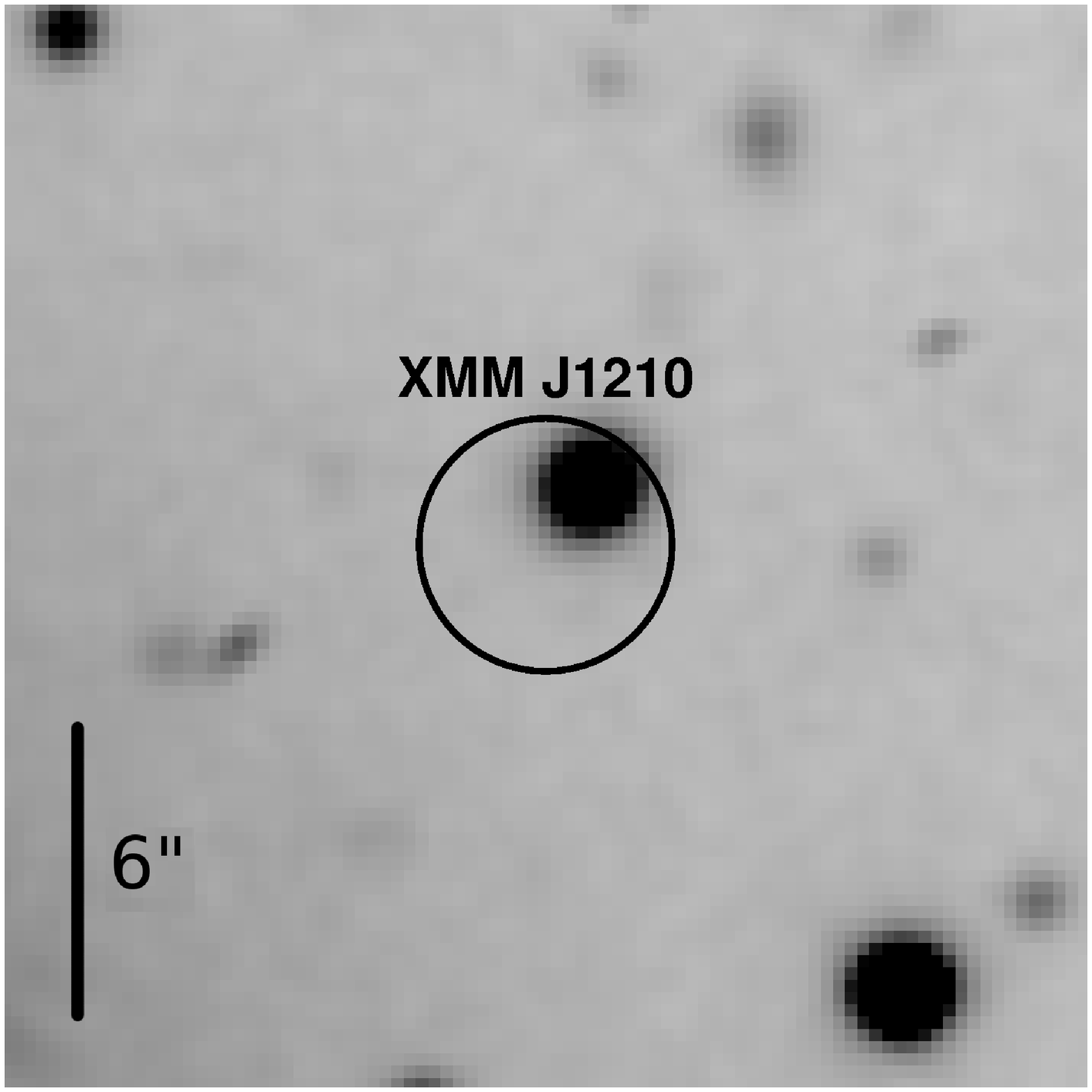}
\includegraphics[width=0.2425\textwidth]{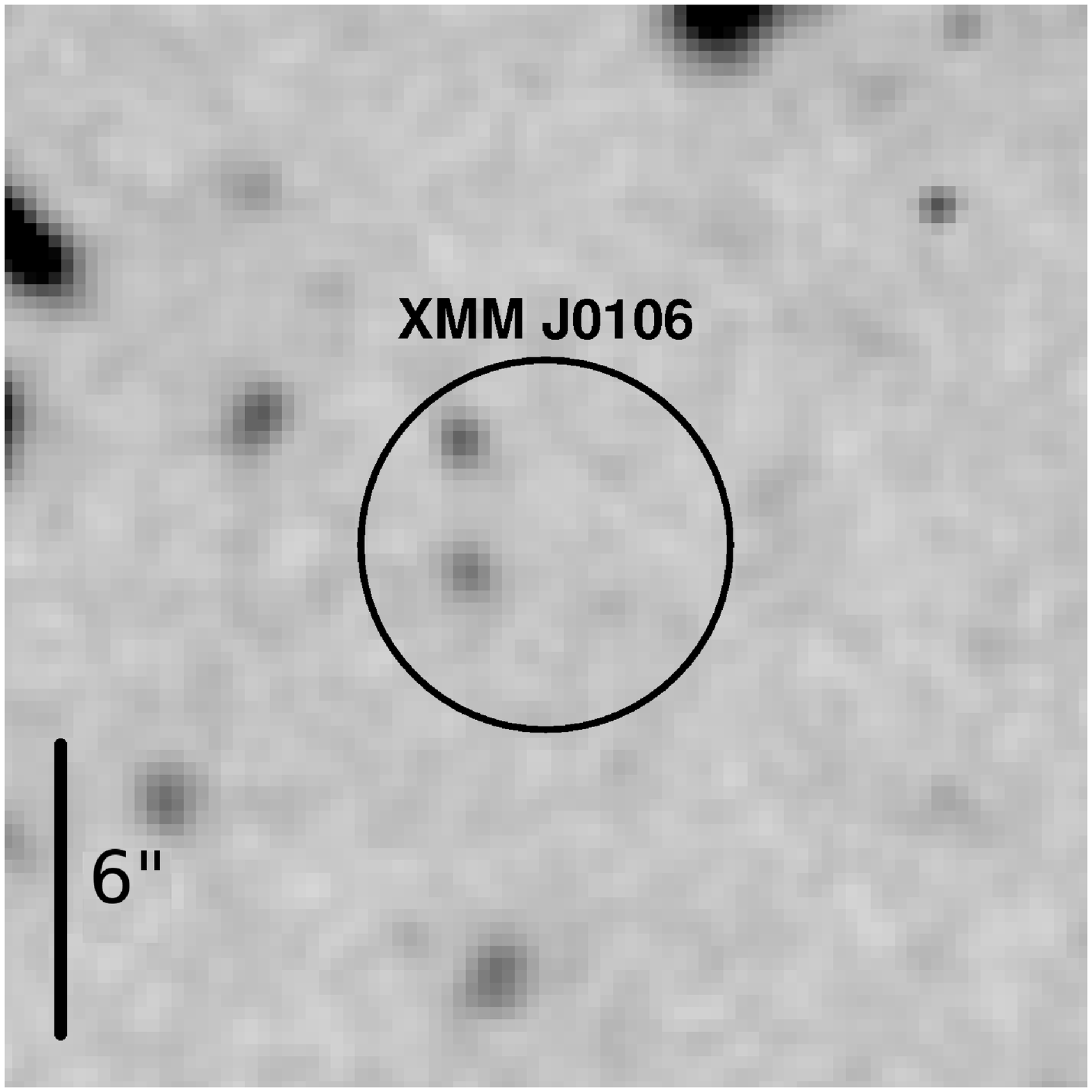}
\includegraphics[width=0.2425\textwidth]{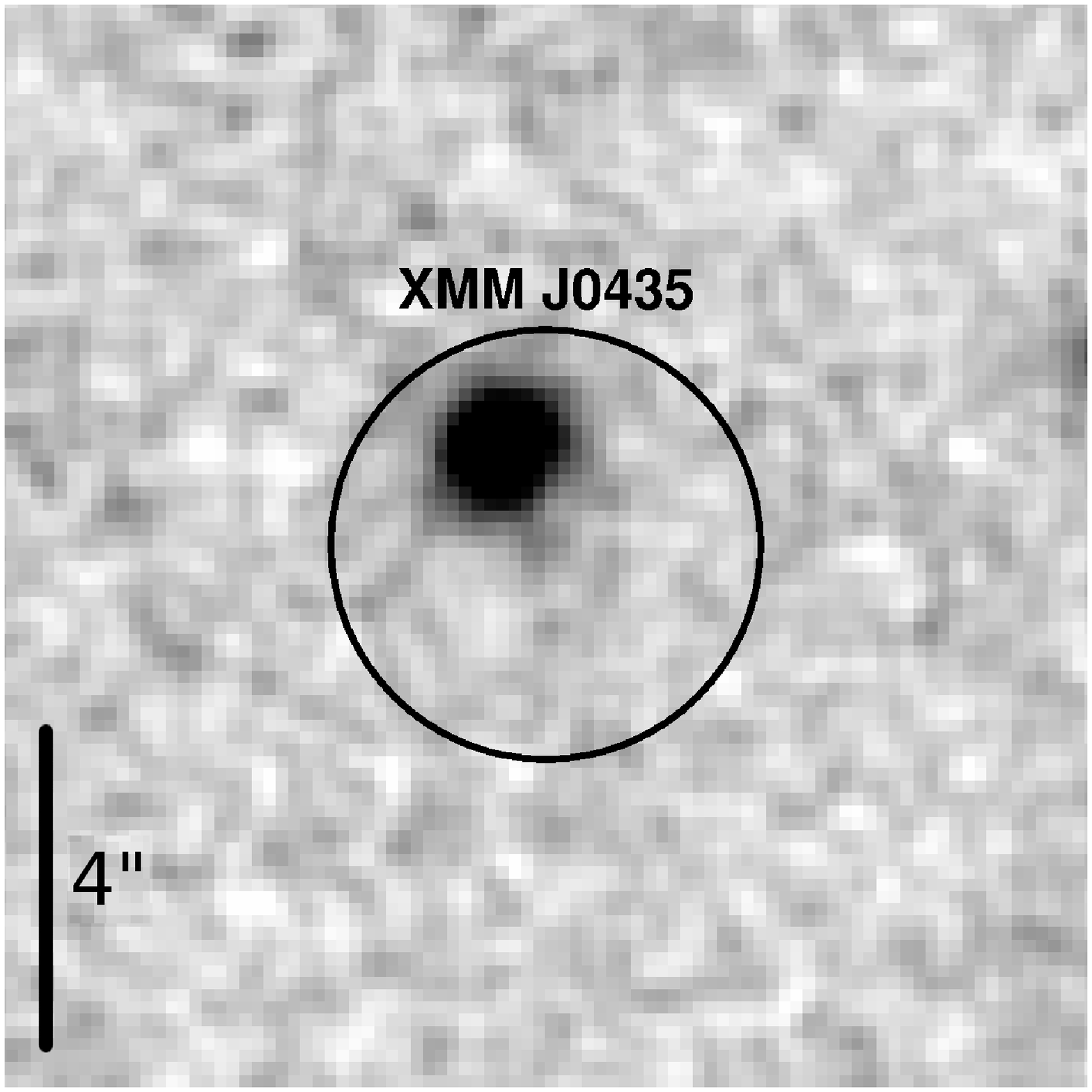}
\vspace{0.3cm}
\includegraphics[width=0.2425\textwidth]{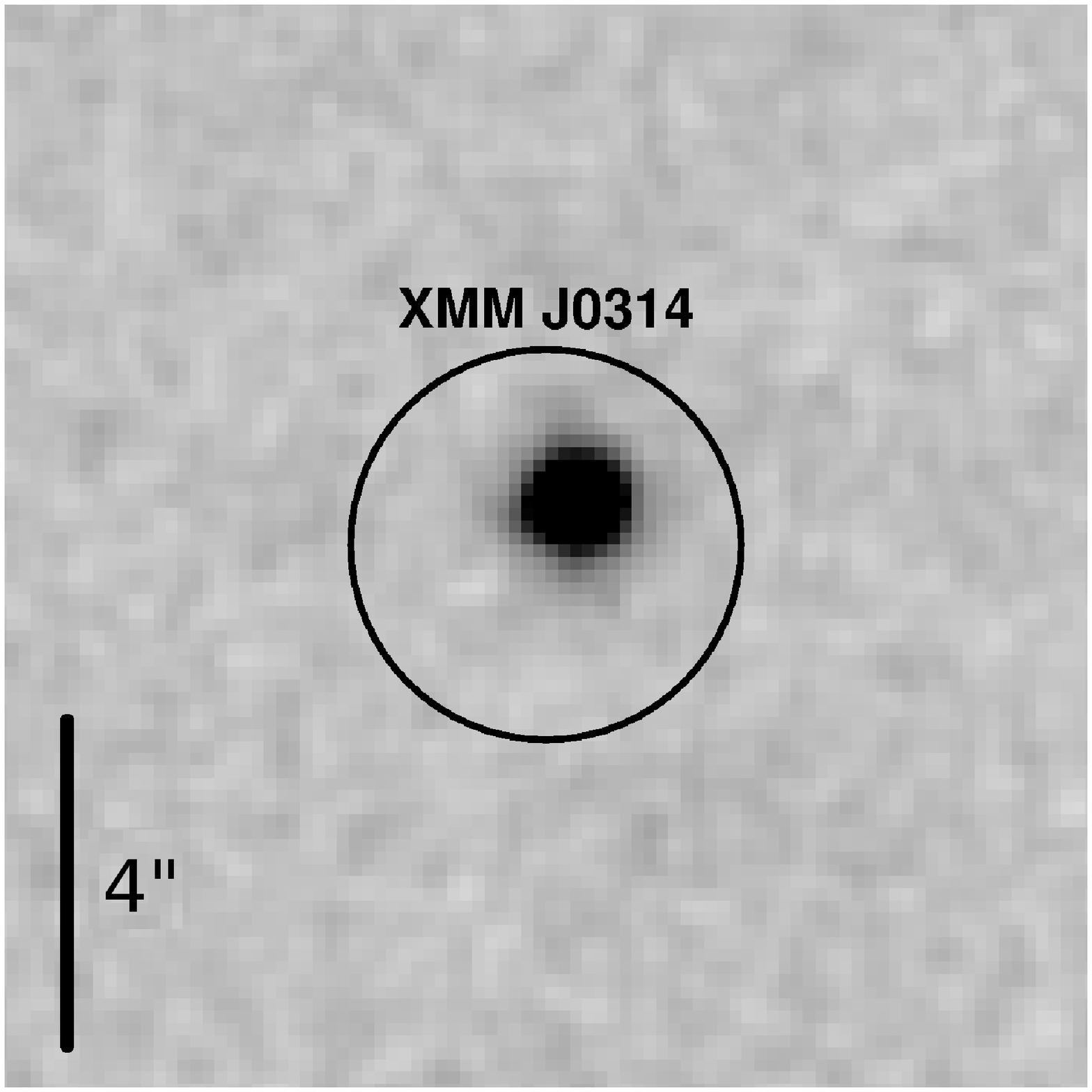}
\includegraphics[width=0.2425\textwidth]{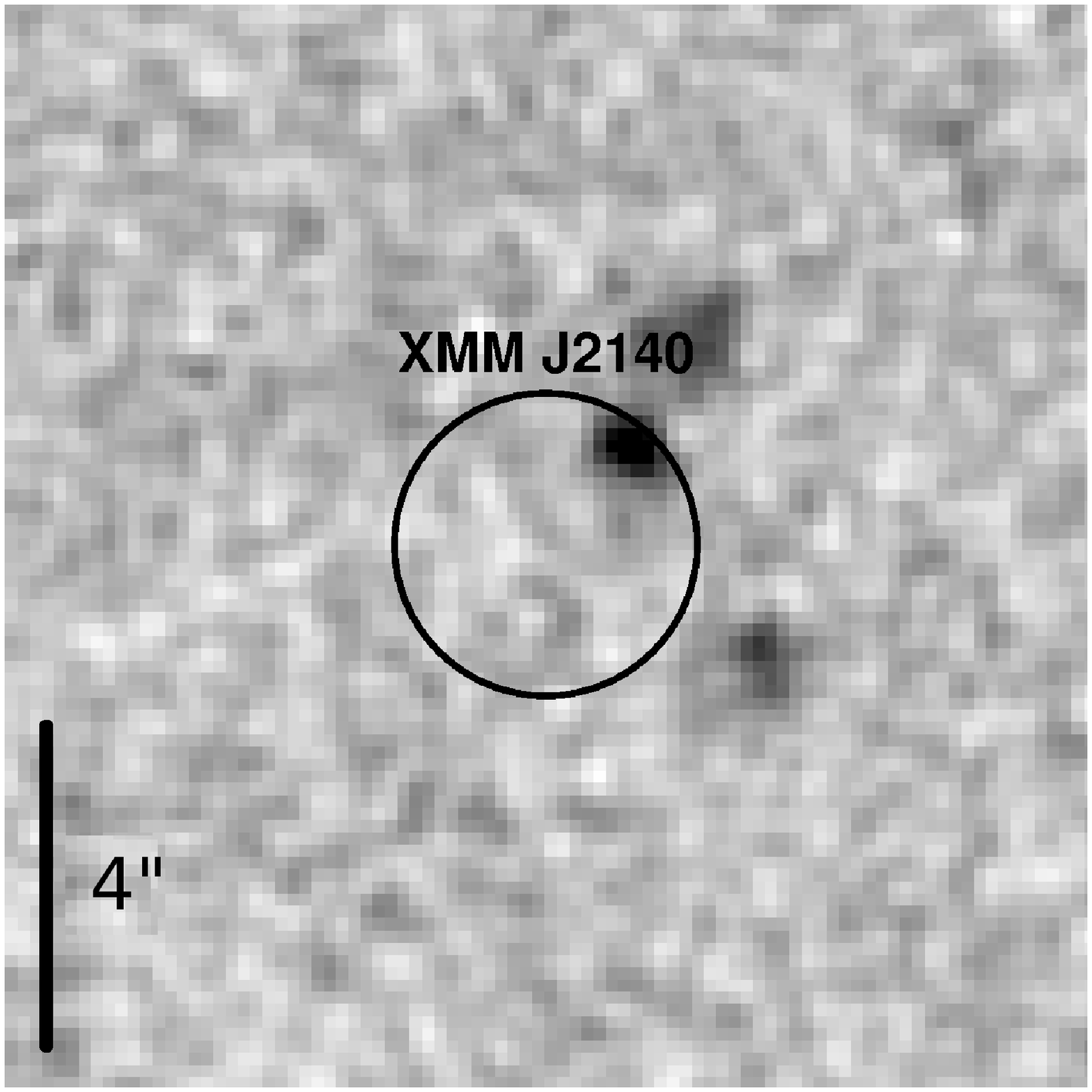}
\includegraphics[width=0.2425\textwidth]{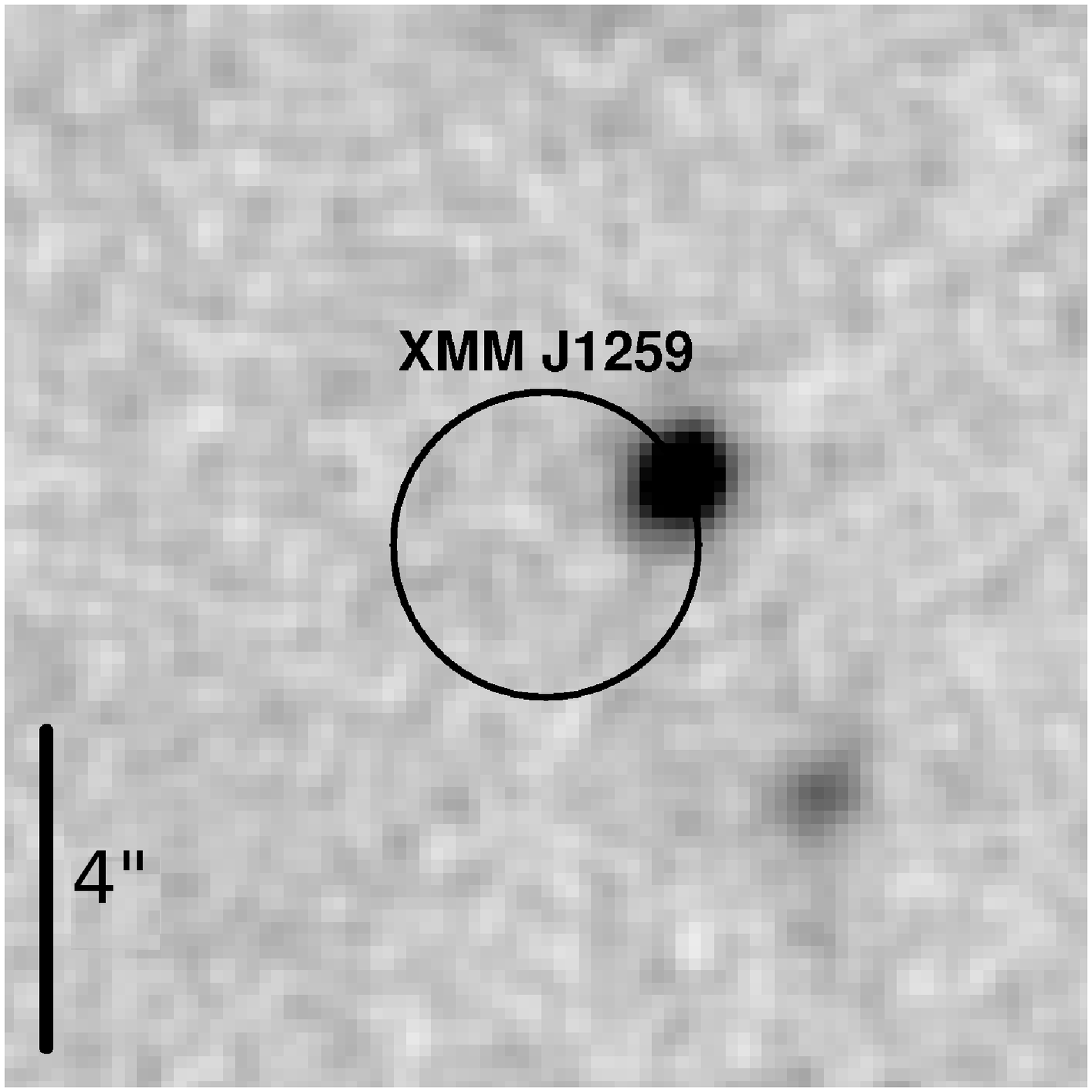}
\includegraphics[width=0.2425\textwidth]{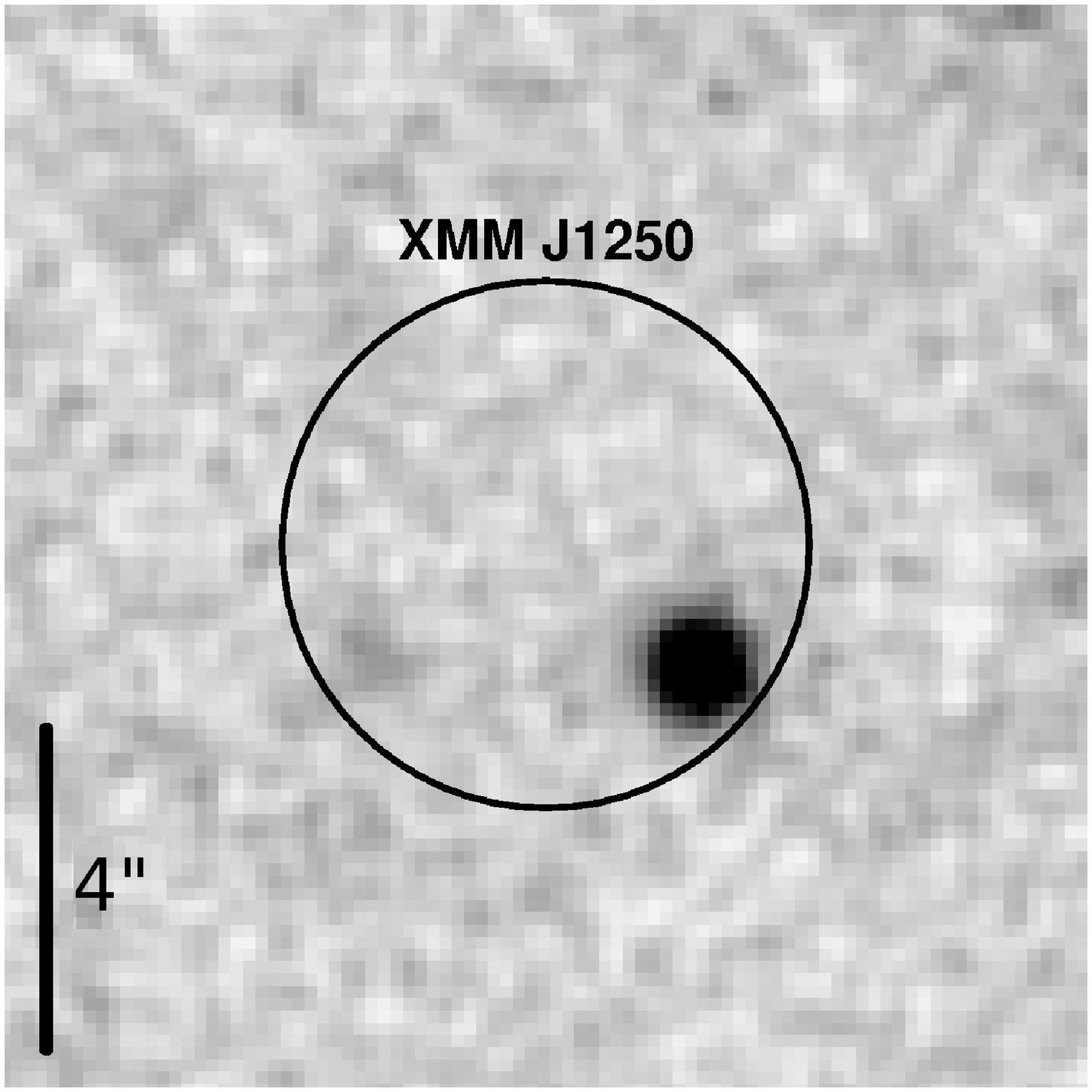}
\caption{R images of the fields of the observed INS candidates. Candidates \candzsf, \candosf\ and \candtoe\ were observed with the ESO-VLT in period P79 while candidates \candtfoe, \candtfie, \candtsf, \candszf\ and \candseo\ were observed using the SOAR telescope. In these images, north and east point upwards and to the left, respectively. Circles show the 90\% confidence level error on the positions.\label{fig_FC}}
\end{figure*}
% -----------------------------------------------------------------------------------------------------------

\subsubsection{Photometry}
Two different approaches were adopted to measure magnitudes: standard PSF fitting, as implemented in the \textsf{DAOPHOT} \citep{ste87} package for the IRAF environment and, for the fields of candidates located at high Galactic latitudes ($|b|\ga20^{\circ}$, see Table~\ref{tab_cand}), variable elliptic apertures and background maps using SExtractor 2.5.0\footnote{\texttt{http://terapix.iap.fr/rubrique.php?id\_rubrique=91}}. 

% ----- TABLE - OPTICAL COUNTERPARTS ------------------------------------------------------------------------
\begin{table*}
\begin{minipage}[t]{\textwidth}
\caption{Results of follow-up investigations\label{tab_resmag}}
\centering
\renewcommand{\footnoterule}{}
\begin{tabular}{l c l l l c c c c}
\hline \hline
Candidate & $r$ & $R$ & $U-B$ & $B-R$ & $LR$ & $P_{\rm{chance}}$ & $e$ & $\log(f_{\rm{X}}/f_{\rm{R}})$\footnote{The X-ray flux is computed in the 0.15\,--\,3\,keV energy band assuming an absorbed power-law model.}\\ 
 & (arcsec) &  &  &  &  & (\%) & & \\
\hline
\candosf  & 1.55 & 20.13(3)   & ?	 & 1.35(3)    & 8.4  & 2.8  & 0.037 & 0.18 \\
\candtoe  & 2.75 & 24.51(9)   & ?	 & $\ga$\,2.0 & 1.2  & 22.4 & 0.178 & 1.93 \\
          & 1.68 & 24.58(9)   & ?	 & $\ga$\,1.9 & 1.2  & 22.4 & 0.162 & 1.96 \\
\candtfoe & 1.20 & 22.15(4)   & -0.51(8) & 0.93(7)    & 15.2 & 2.3  & 0.068 & 0.77 \\
\candtfie & 0.67 & 21.37(3)   & -0.13(7) & 1.43(4)    & 32.1 & 1.7  & 0.022 & 0.41 \\
\candtsf  & 1.26 & 23.78(9)   & -1.08(16)& 1.06(15)   & 6.0  & 6.4  & 0.156 & 1.40 \\
\candszf  & 1.34 & 21.64(4)   & -1.08(4) & 0.76(5)    & 25.5 & 1.0  & 0.098 & 0.61 \\
\candseo  & 1.80 & 22.19(4)   & -0.71(4) & 0.74(4)    & 8.4  & 4.1  & 0.029 & 0.57 \\
\hline
\end{tabular}
\end{minipage}
\end{table*}
% -----------------------------------------------------------------------------------------------------------

PSF fitting using a two-dimensional Gaussian function was adopted whenever there were a fair number of isolated and bright non-saturated stars available, allowing a good determination of the modeled PSF to apply to all detected sources on the image (including faint and overlapping stars). We note, however, that none of the fields is severely crowded, not even the one located at $b\sim0^{\circ}$. When this was not possible, we adopted SExtractor instead. This software is most commonly used to reduce galaxy survey data but it also performs well in moderately crowded stellar fields. As many extragalactic objects were likely to be present in our high Galactic latitude fields, its usage is convenient since it is possible to derive, for instance, information on how elongated an object is; additionally, making use of a neural network, SExtrator classifies any given object as stellar or non-stellar. However, this classification is less reliable for faint fluxes. Its main advantage is that flux measurements of extended objects tend to be more accurate when compared to the ones obtained using fixed aperture photometry, since elliptical apertures with variable sizes (based on the object intensity) are used instead. Another advantage is that a global background map is created, which better accounts for local spatial variations of brightness due e.g. to nebular emission or caused by scattered light from bright objects. 

Using SExtractor, we created catalogues gathering position, intensity and shape information on every optical object present in our high Galactic latitude frames. Source detection was carried out by convolving a Gaussian filter having the mean image FWHM. The resulting list of sources in each field was then correlated to keep the optical objects which were detected in all filters. In the following subsections we use the information extracted from these optical catalogues to discuss the nature of our sample of INS candidates and field objects. 

\subsubsection{X-ray/optical associations\label{sec_Xopt}}
We found at least one optical object inside the X-ray error circles (90\% confidence level) of all INS candidates but one, the X-ray brightest source \candzsf\ (see Fig.~\ref{fig_FC}). In this case no optical object brighter than the limiting magnitude of our present data is present within $\sim$\,$4.3''$ ($\ga$\,5\,$\sigma$) from the position of the X-ray source (see discussion in \ref{sec_limmag}). 

% ----- FIGURE - HISTOGRAMS LR ------------------------------------------------------------------------------
\begin{figure*}
\centering
\includegraphics[width=0.2425\textwidth]{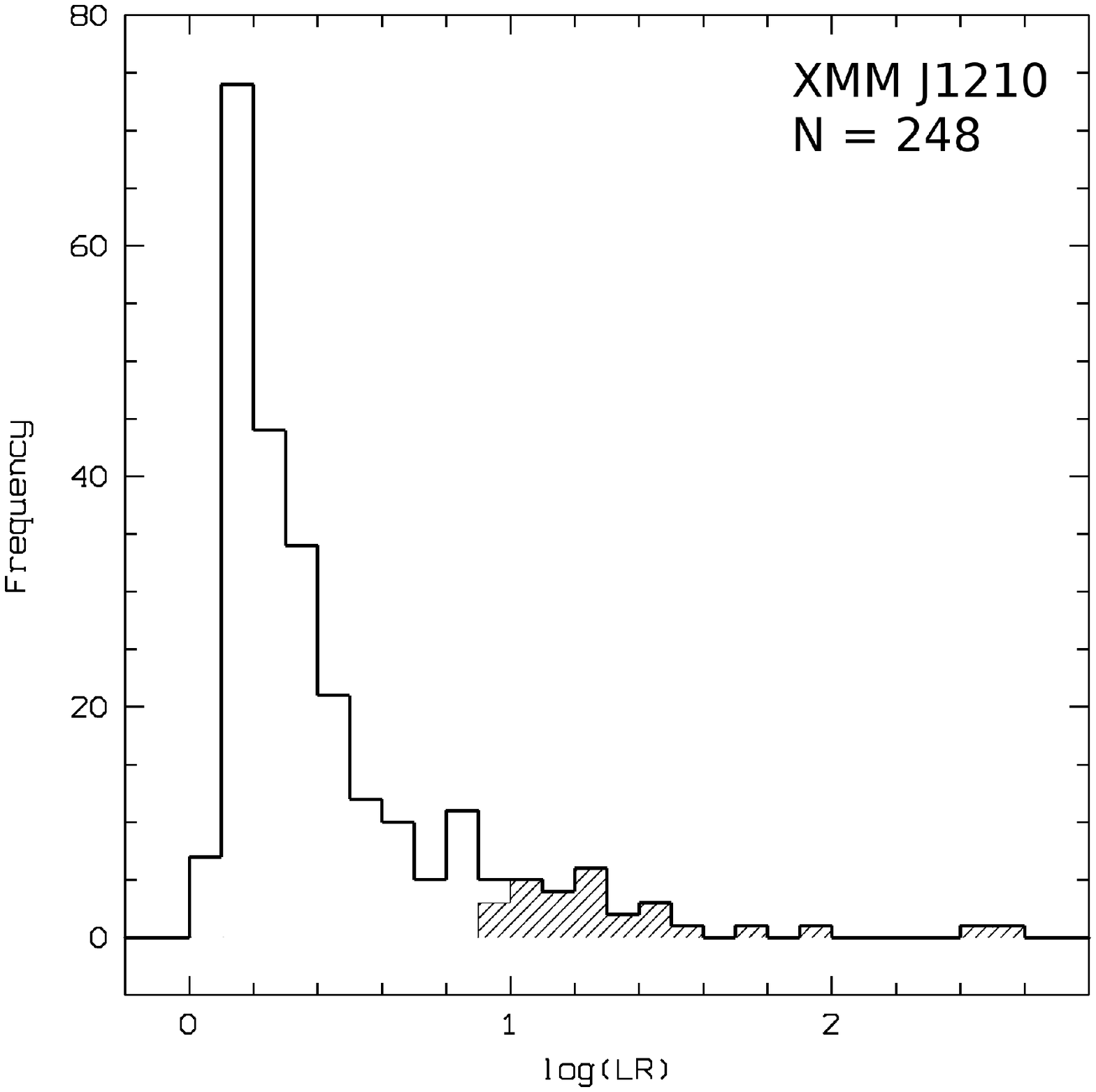}
\includegraphics[width=0.2425\textwidth]{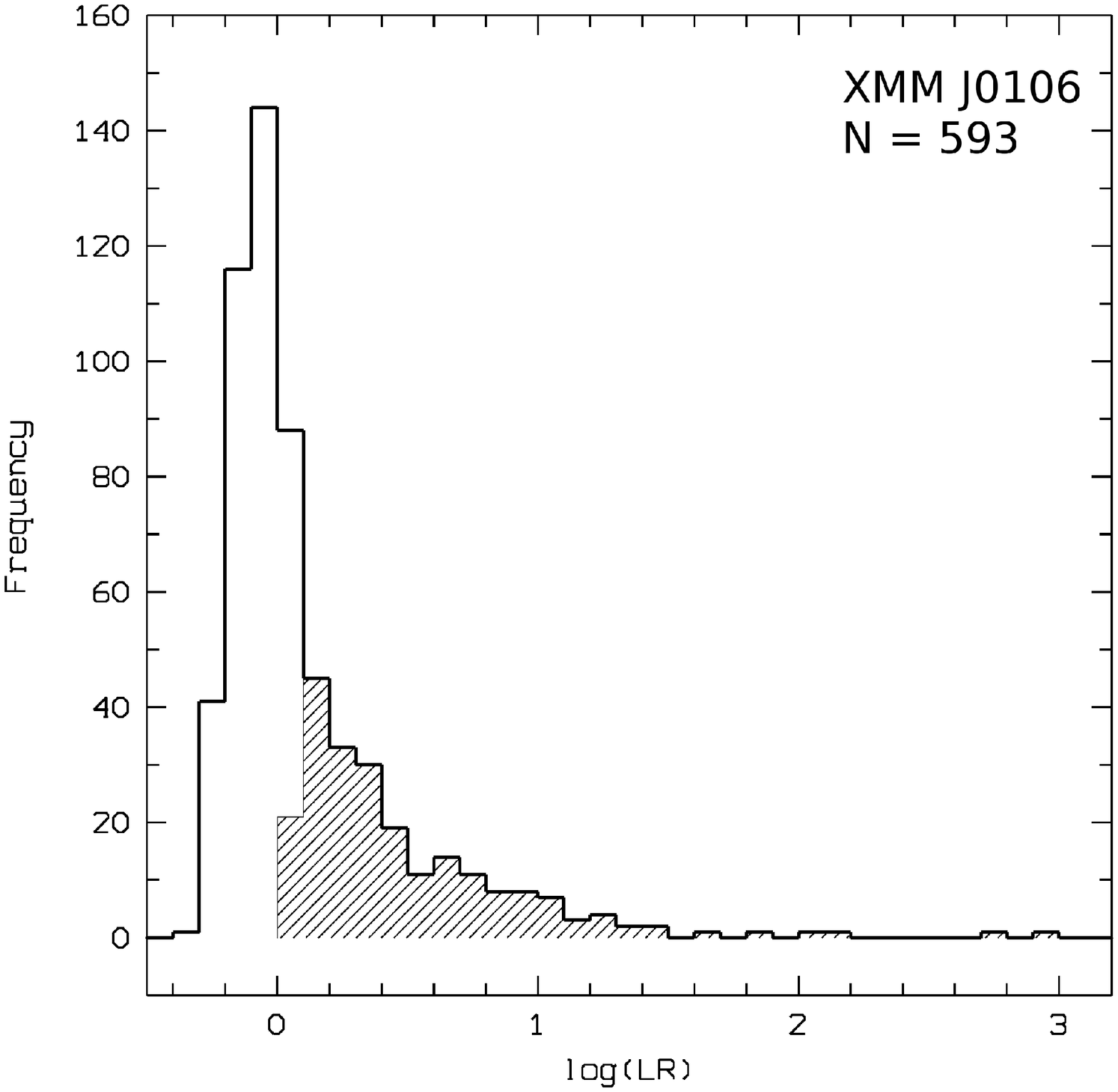}
\includegraphics[width=0.2425\textwidth]{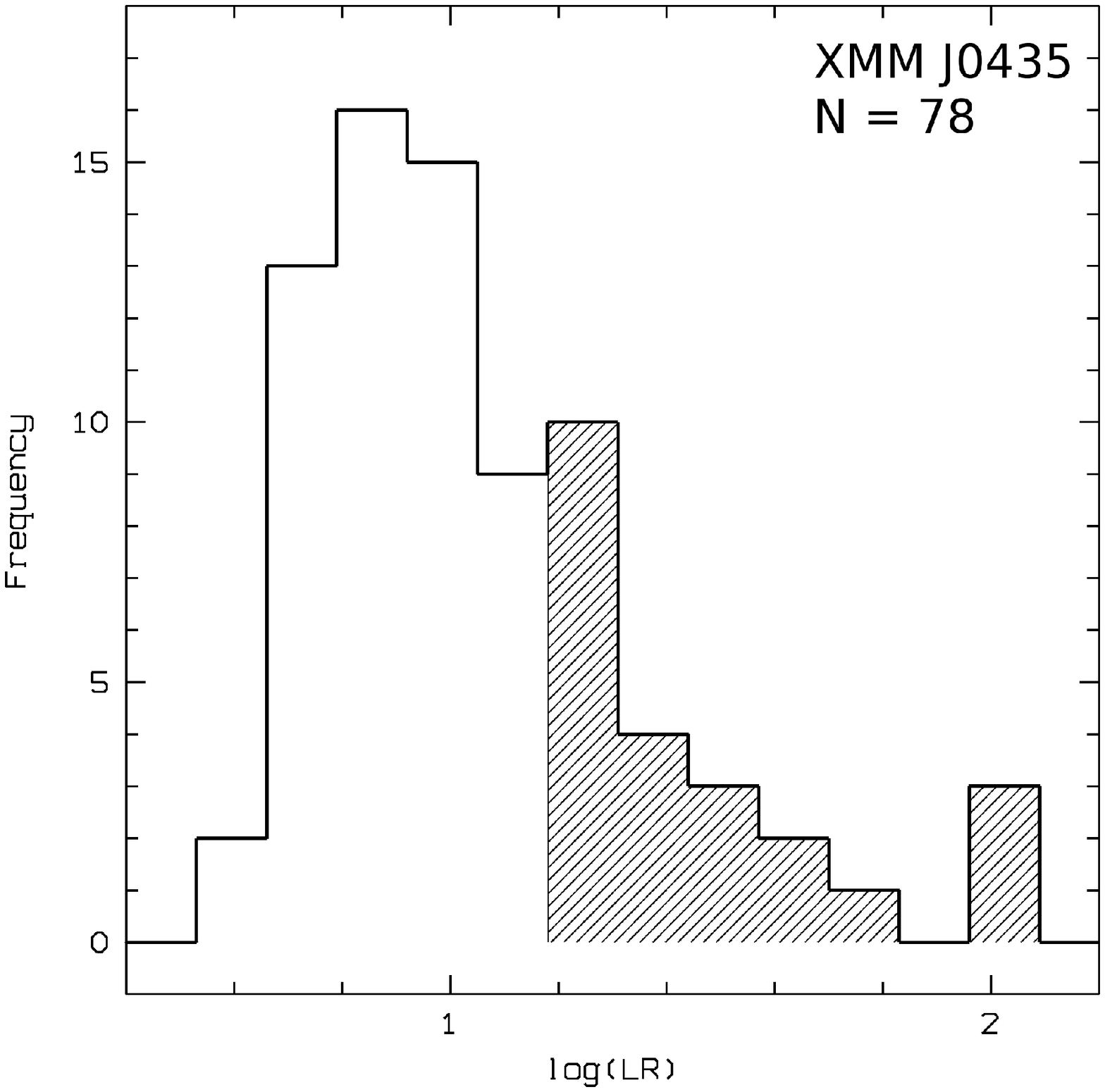}
\includegraphics[width=0.2425\textwidth]{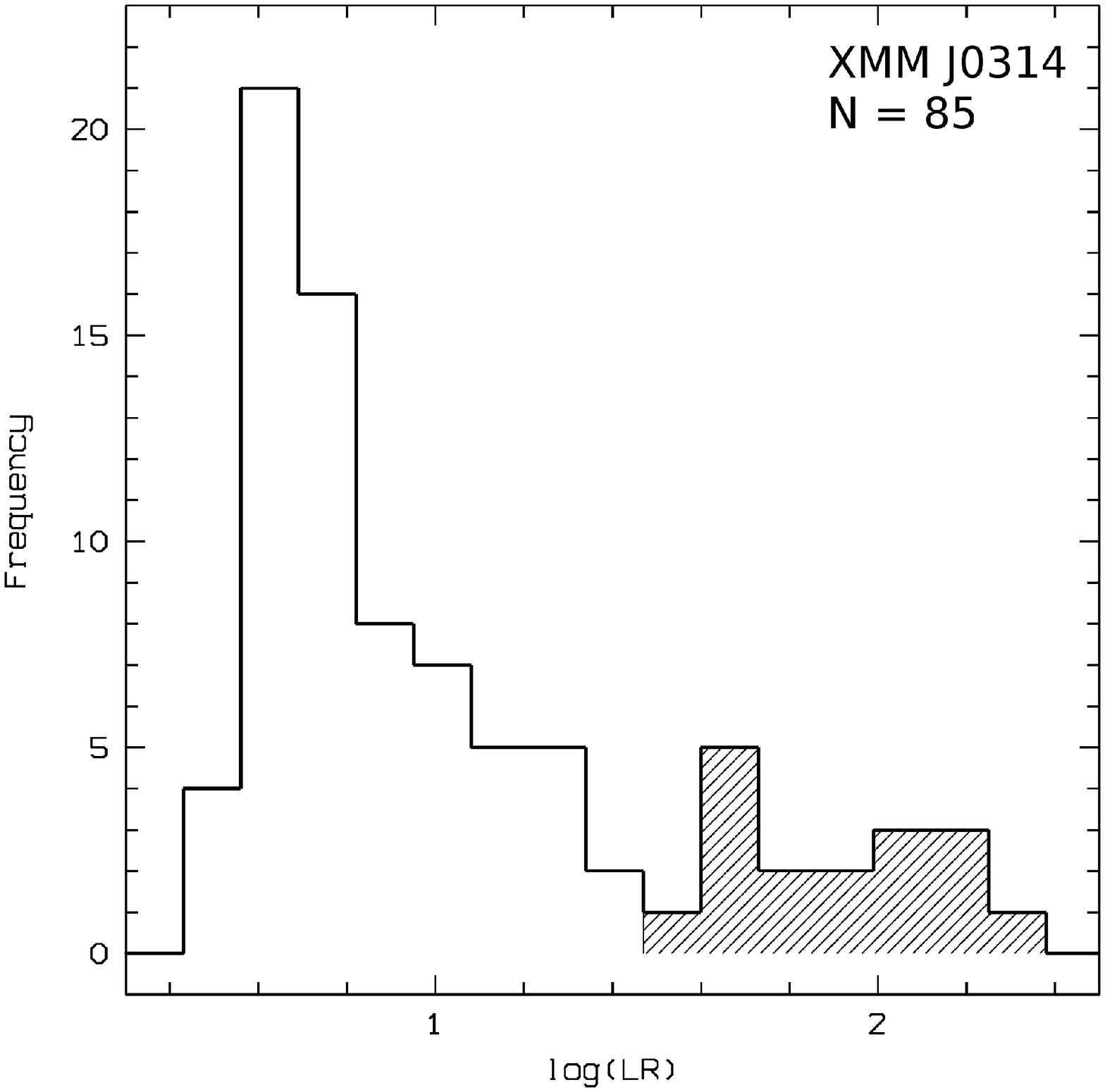}
\\
\vspace{0.5cm}
\includegraphics[width=0.2425\textwidth]{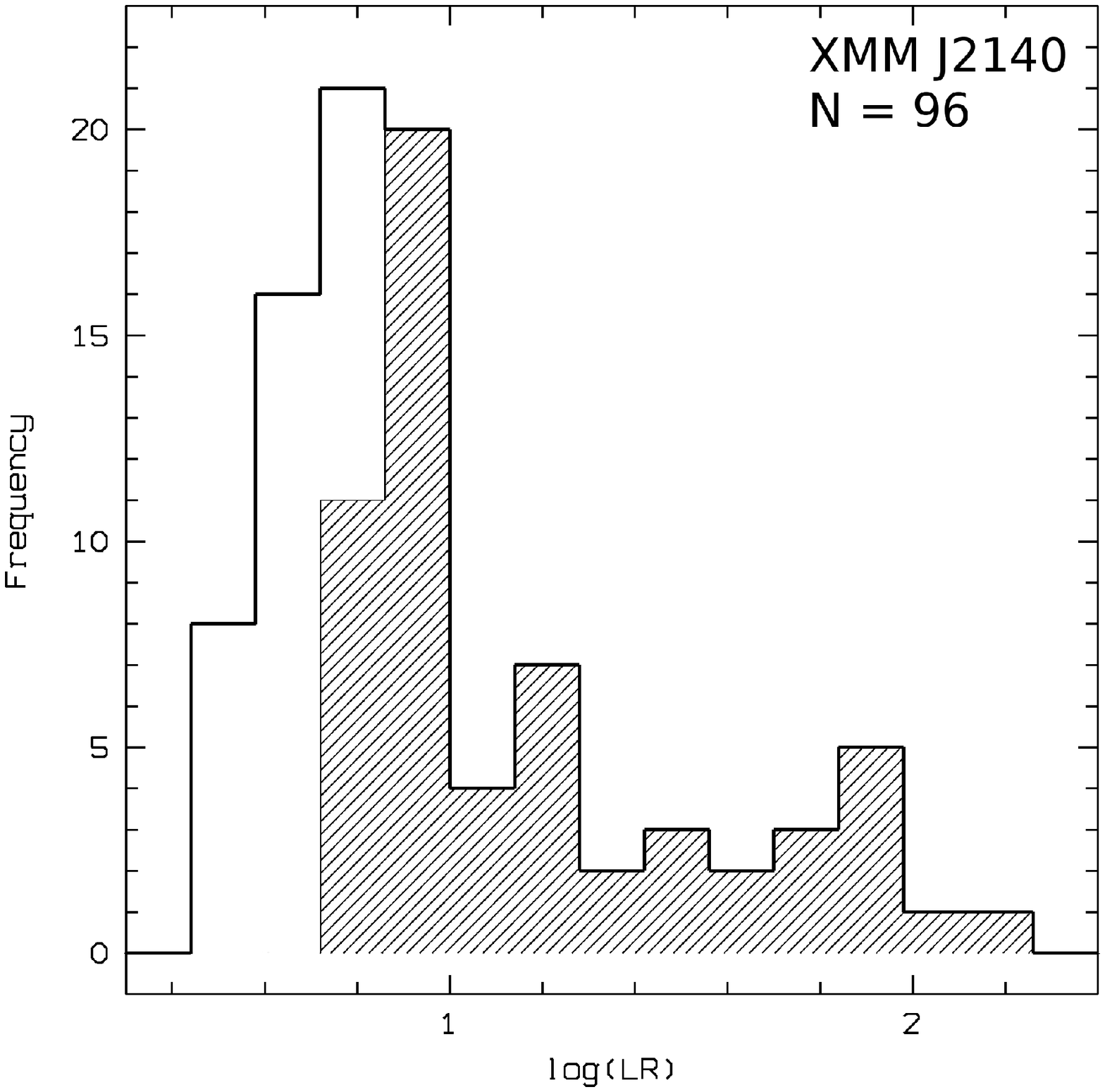}
\hspace{0.5cm}
\includegraphics[width=0.2425\textwidth]{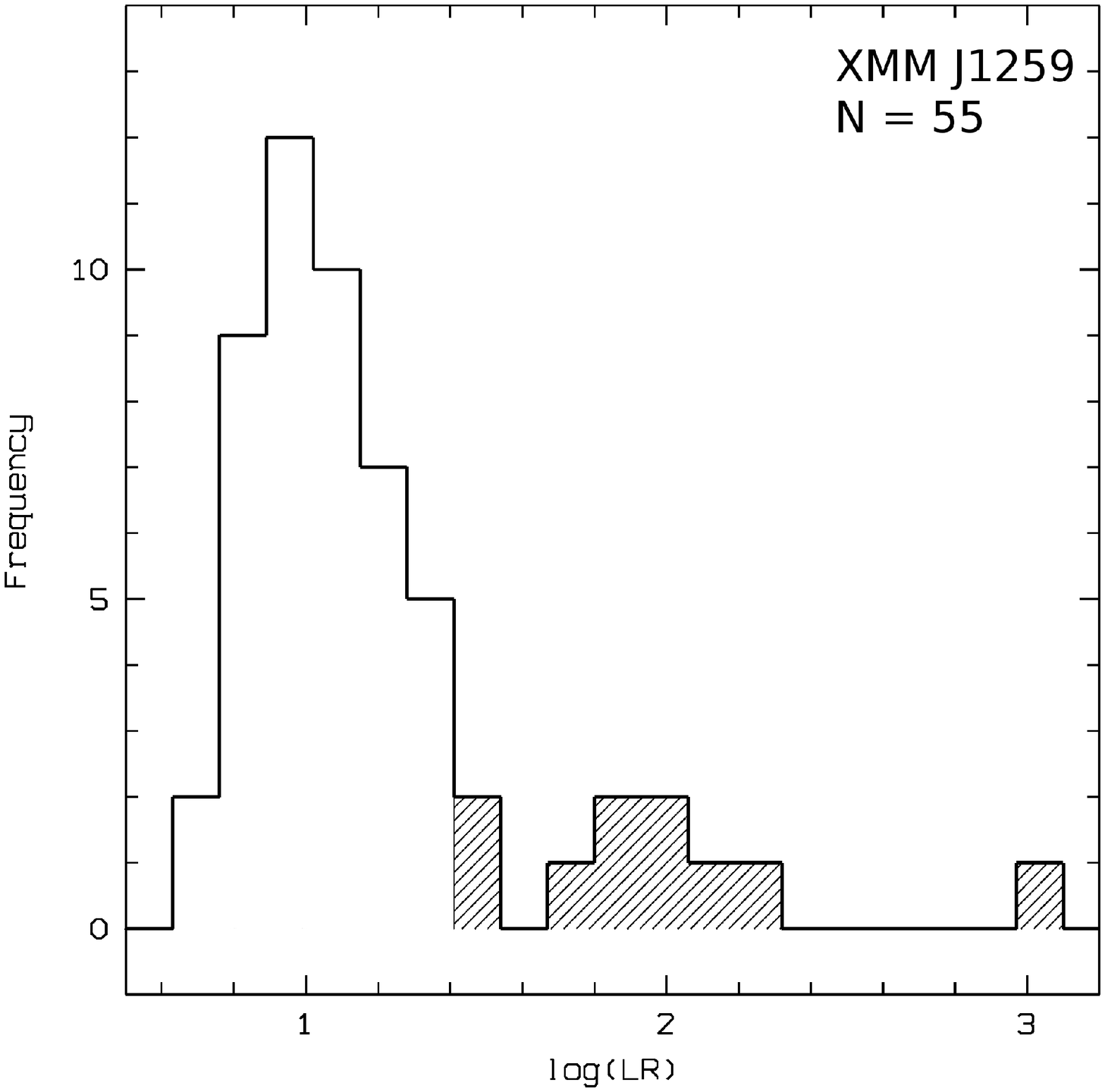}
\hspace{0.5cm}
\includegraphics[width=0.2425\textwidth]{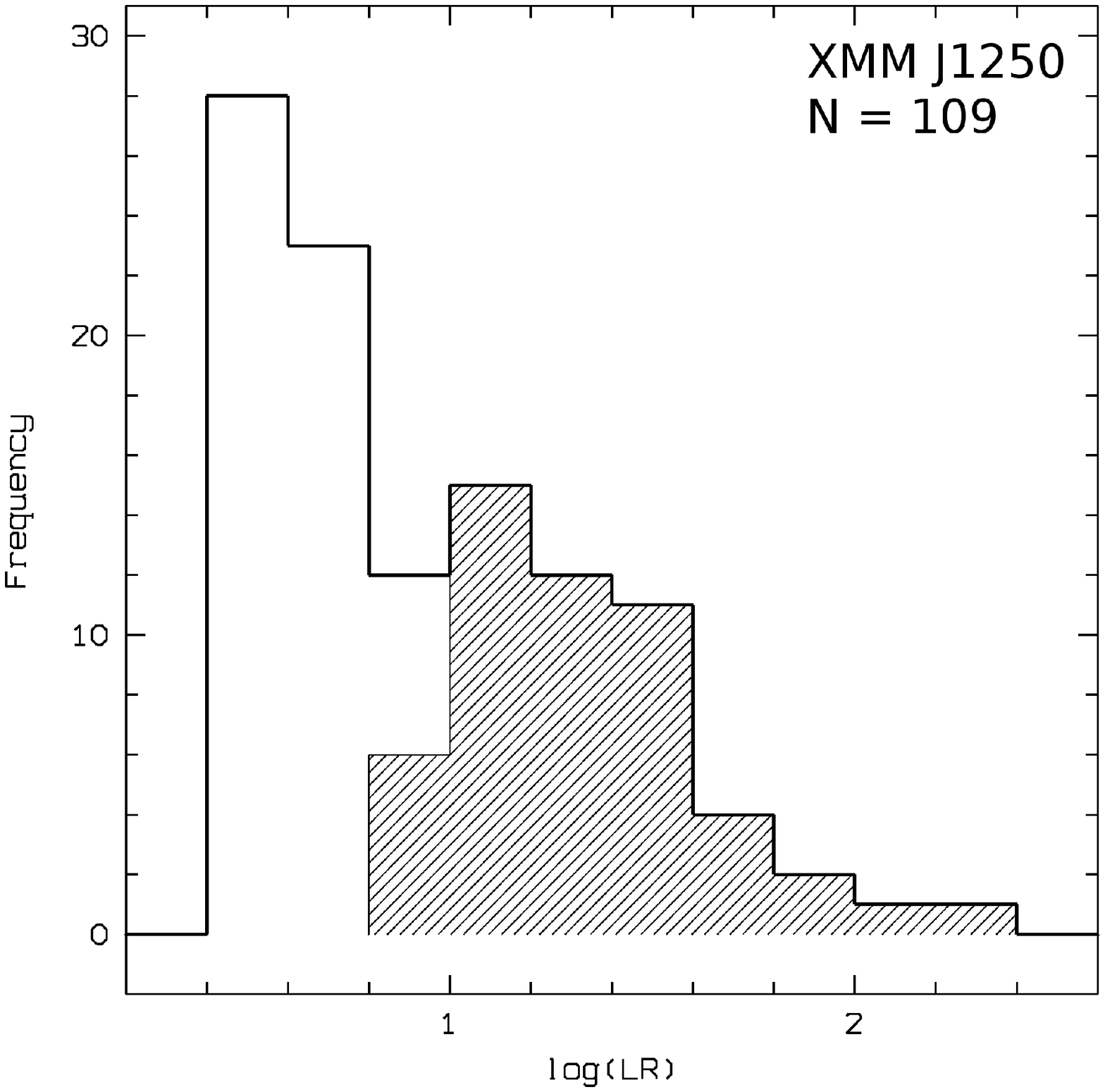}
\caption{Histograms showing the $LR$ distribution of X-ray/optical associations in the fields of candidates for which possible optical counterparts were found. Each histogram is computed from an original number of 1000 Monte Carlo simulations. The number of matchings is given in the right upper corners. Hashed bars show simulated X-ray/optical associations which have a $LR$ greater than or equal to the one found for the X-ray source.\label{fig_histLR}}
\end{figure*}
% -----------------------------------------------------------------------------------------------------------

Table~\ref{tab_resmag} lists the $R$ magnitudes and $B-R$ and $U-B$\footnote{Only for SOAR data.} colours of the possible optical counterparts of the remaining seven sources, as well as their positional offsets relative to the X-ray coordinates and the implied logarithmic X-ray-to-optical flux ratios, uncorrected for absorption.
The candidate counterparts were well detected in all (UBR or BR) filters, with the exception of the two optical objects possibly associated with \candtoe, only detected on the R image. Among the possible counterparts, these were also the faintest ones ($R\sim24.5$).

To estimate the reliability of each X-ray/optical association, we defined a likelihood ratio ($LR$) based on the probabilities that, given the positions, an optical object is the true counterpart of the X-ray source or is just a random object that happens to lie angularly close to its position. 
Following \cite{rui77}, $LR$ can be defined as:

\begin{equation}\label{eq_lr}
LR = \frac{1}{2\lambda}\exp\Bigg[\frac{x^{2}}{2}\Big(2\lambda - 1\Big)\Bigg]
\end{equation}
where $\lambda$ is defined as $\lambda = \pi\sigma^2\rho$, $\sigma$ is the error on the X-ray and optical positions and $\rho$ is the surface density of brighter or equally bright objects on the image. The dimensionless variable $x = r/\sigma$ is the significance of the position offset ($r$).

We list in Table~\ref{tab_resmag} the $LR$ quantities for every possible X-ray/optical association, which ranges from $\sim$\,1 to 32. Although objects with large $LR$ are more likely to constitute a true association, the computation of actual probabilities of identification requires a calibration of $LR$ taking into account the \emph{a priori} probability that any X-ray source has a counterpart in the optical sample considered (see e.g. \citealt{pin08}). 

$LR$ strongly depends on the local density of optical objects; the chance of a spurious association rising with the number of objects of similar or brighter magnitudes. In order to derive the $LR$ distribution of spurious associations and therefore estimate the chance probability of X-ray/optical associations in our data, we performed Monte Carlo simulations randomly changing the position of the X-ray source accross the optical fields. For a given simulated X-ray source, we then computed the $LR$ quantity given by Equation~(\ref{eq_lr}) for every optical object (if any) lying inside $r_{90}$. The overall distribution of positive matches (among 1000 simulations) can be seen in the histograms of Fig.~\ref{fig_histLR}.
As expected, the distribution of $LR$ is highly asymmetrical, strongly peaking at low values and then monotonically extending towards high $LR$. Hashed bars in Fig.~\ref{fig_histLR} highlight the number of simulated random X-ray/optical associations with $LR_i\ge LR_{\rm X}$, where $LR_i$ and $LR_{\rm X}$ are the likelihood ratios of simulated and the actual X-ray/optical association of the given field.

In general, the simulations show that the probability of a chance association is low for the optical objects found inside $r_{90}$ on the real data (Table~\ref{tab_resmag}). Out of 1000 simulations, we usually found that less than 3\% of the cases had an X-ray/optical association with a likelihood ratio greater than or equal to the real one.
The two exceptions are the associations found for source \candtoe, with $P_{\rm chance}\sim22\%$ and, at a smaller degree, for source \candtsf, with $P_{\rm chance}\sim6\%$. These are the two X-ray sources with the faintest optical candidates and the only ones with $\log(f_{\rm{X}}/f_{\rm{R}})>1$. 
However, whereas the two candidate counterparts for \candtoe\ were detected only in the R band, the one for \candtsf\ has been well detected in the U, B and R filters, and it shows very blue colours. We argue in the following that, based on the results obtained here and together with the analysis of their colours and X-ray-to-optical flux ratios (Section~\ref{sec_optcount}), the optical objects found inside the X-ray error circles of all INS candidates are very likely to constitute their true optical counterparts, with, again, the exception of source \candtoe.

% ----- FIGURE - COLOUR DIAGRAM -----------------------------------------------------------------------------
\begin{figure*}
\centering
\includegraphics[width=0.498\textwidth]{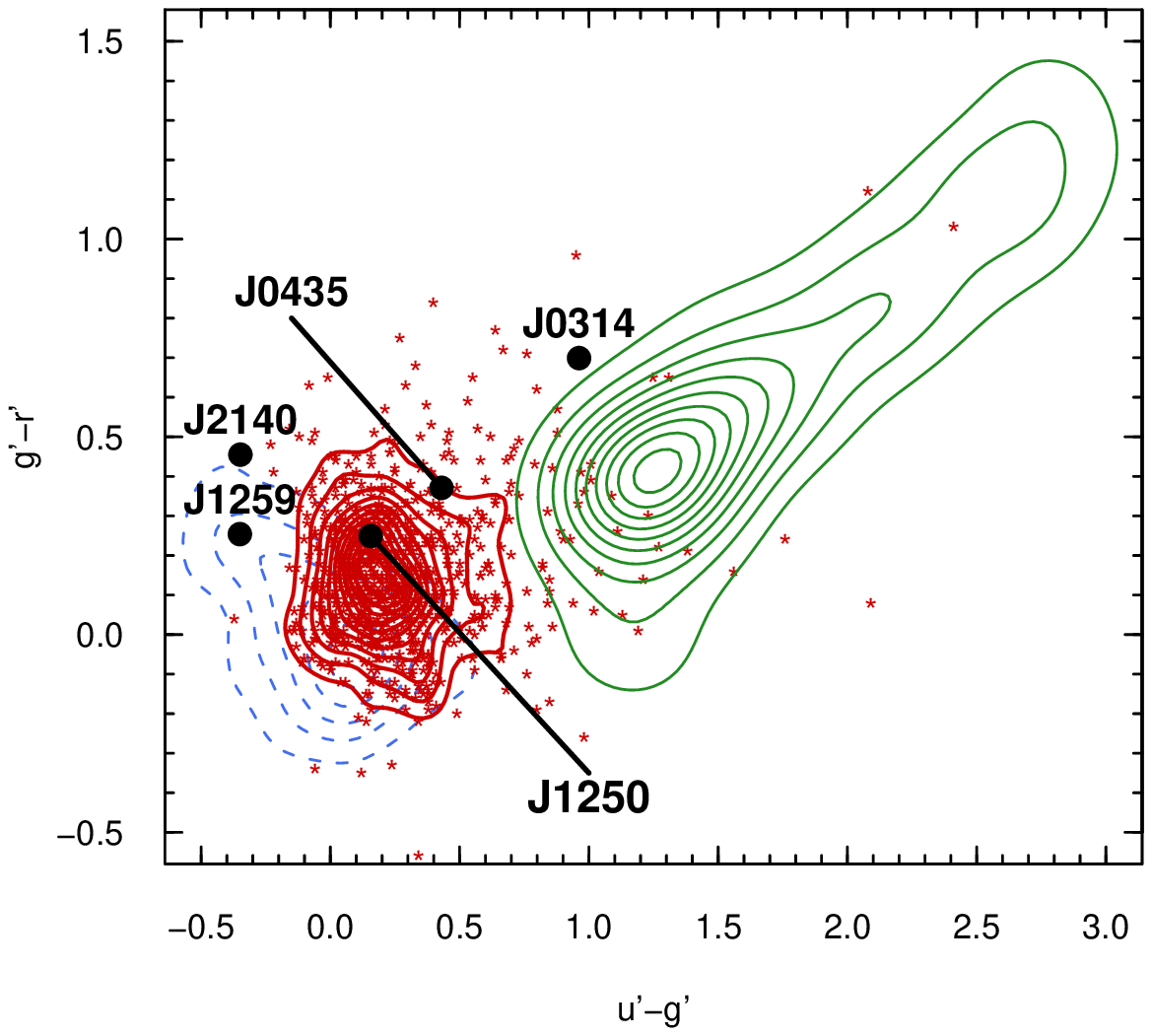}
\includegraphics[width=0.498\textwidth]{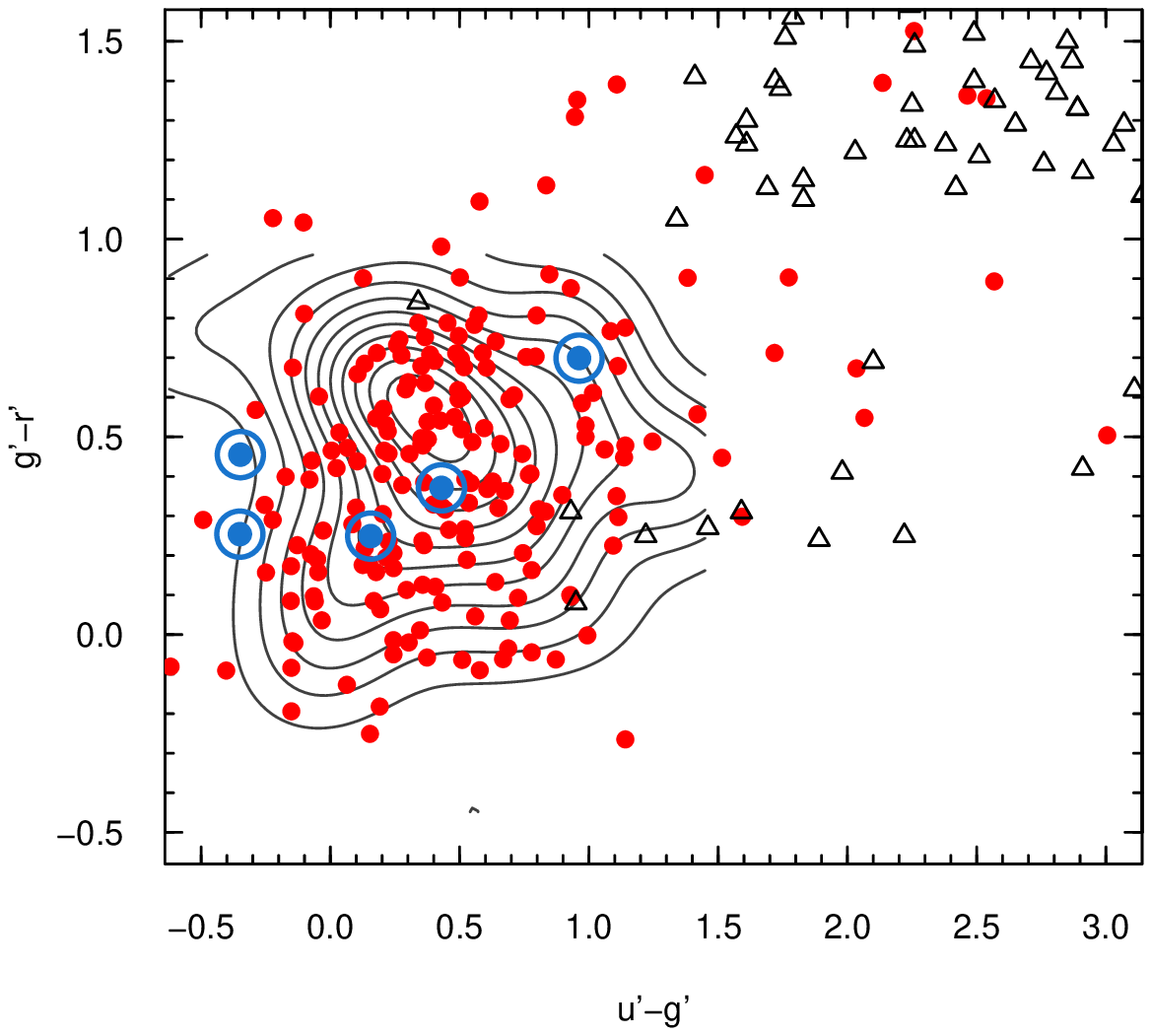}
\caption{\emph{Left:} Colour-colour $g' - r'$\,$\times$\,$u' - g'$ diagram comparing the locations of different astrophysical populations of X-ray emitters. Red stars (and contours) show the position of identified QSOs at redshifts lower than 3. Hashed blue and solid green contours are CVs and late-type stars from the SDSS, respectively (see text). The five INS candidates observed with SOAR are represented by plain black circles with labels. \emph{Right:} Same diagram showing the colours of field objects detected in the three (U, B and R) filters of our SOAR observations (plain red circles). The population of SDSS objects located at a similar column densities is shown for comparison as a density contour. Open triangles are identified quasars located at redshifts greater than 3. Optical candidates of the soft X-ray sources are shown as large blue circles.\label{fig_coldiag}}
\end{figure*}
% -----------------------------------------------------------------------------------------------------------
\subsubsection{Identification with other classes of X-ray emitters\label{sec_optcount}}
As mentioned before, some populations of astrophysical objects are expected to pollute our sample of INS candidates. 
In order to identify these among the sources having optical candidates, we compared their positions in the colour-colour $g' - r'$\,$\times$\,$u'-b'$ diagram with those of the major classes of X-ray emitters (left pannel of Fig.~\ref{fig_coldiag}). The magnitudes in the Sloan photometric system were computed using the Johnson-Morgan-Cousins transformation equations in \cite{fuk96}. In this diagram are plotted, as contours, the spectroscopically identified population of quasars \citep{ric01} and cataclysmic variables \citep[and references therein]{dil08} from the SDSS, as well as late-type stars \citep{bil05}. Only the five INS candidates observed with SOAR are shown, since we do not have $U$ magnitudes for the two candidates observed with the ESO-VLT (sources \candosf\ and \candtoe). We note, however, that the $B-R$ colours of the optical candidates of these sources stand among the ``reddest'' of our sample (Table~\ref{tab_resmag}).

The $u' - g'$ and $g' - r'$ colours of the optical candidates of the SOAR targets are clearly consistent with those of AGN and CVs. Among the five sources, the proposed counterpart of \candtfie\ is the only one which stands somewhat more isolated from the bulk of these two populations of X-ray emitters, although some quasars at still low redshifts from \cite{ric01} do show similar colours. None of the optical sources exhibits colours typical of late-type stars. The strong blue/UV excess, with $u'-g'<0$, shown by the proposed counterparts of sources \candszf\ and \candtsf\ is remarkable. The colours of the optical objects in the error circles of \candseo\ and \candtfoe\ are somewhat more intermediate but still consistent with those of AGN and CVs.

Interestingly, when compared to other uncatalogued field objects present in our SOAR frames and \emph{detected in all filters}, the colours and even the UV excess exhibited by our sample of optical candidates do not stand out from the overall field population (see right pannel of Fig.~\ref{fig_coldiag}). For further comparison, we also plot as density contours SDSS objects present in five regions of the sky located at similar column densities as the SOAR fields. More precisely, we gathered SDSS entries in a SOAR field-of-view centered at five different directions, requiring $r'$ magnitudes between 21 and 24, which is the range covered by our SOAR field objects, and $u'$, $g'$ and $r'$ errors smaller than 0.5. Overall, the plot shows that the SOAR and SDSS objects constitute the same population of ``blue field objects'' located at high Galactic latitudes, with colours not particularly distinguishable from those of the sample of soft X-ray counterparts. Apart from these blue objects, a smaller fraction of the SOAR field population occupies the same location in the colour diagram as the one of quasars identified at more remote redshifts ($z>3$; see open triangles in Fig.~\ref{fig_coldiag}). Most ($\ga$\,99\%) of the SOAR objects do not have a correlation with the 2XMMi and none with the FIRST (radio) catalogues. According to \cite{mat08}, the expected density of soft (0.5\,--\,2\,keV) X-ray sources at high Galactic latitudes $|b|>20^{\circ}$, with fluxes greater than $f_{\rm X}>10^{-14}$\,erg\,s$^{-1}$\,cm$^{-2}$ is $\sim$\,100\,deg$^{-2}$, corresponding to $\sim$\,0.75 sources in a typical SOAR field-of-view -- which is roughly in agreement with our results. In summary, despite the blue colours of the field SOAR objects are similar to the ones of the INS candidates, they are not active in X-rays or radio wavelengths.

% ----- TABLE - LIMITING MAGNITUDES -------------------------------------------------------------------------
\begin{table}
\begin{center}
\caption{Upper limits on the $B$, $V$ and $R$ magnitudes for the optical counterpart of \candzsf\label{tab_maglim}}
\begin{tabular}{l c c c c}
\hline\hline
 Filter & $m_{\rm syn}$ & $\sigma_{\rm det}$ & $m_{\rm det}$ & $\log(F_{\rm{X}}/F_{\lambda})$ \\
\hline
 B & 26.0 & 2.4 & 26.2(5) &  $\ga$\,2.2\\ 
 V & 25.5 & 2.5 & 25.9(4) &  $\ga$\,3.1\\
 R & 24.2 & 2.4 & 25.5(5) &  $\ga$\,3.2\\
\hline
\end{tabular}
\end{center}
\end{table}
% -----------------------------------------------------------------------------------------------------------

We analysed the extent distribution of field objects and optical counterparts using the ellipticity derived by SExtractor. The goal was to verify the possibility of separating stars/AGN from galaxies by measuring their point spread functions. Defining ellipticity as usual ($e = 1 - b/a$, where $a$ and $b$ are the semi-major and semi-minor axis lengths, respectively) we found that most ($\ga$\,60\%) of the SOAR field objects are not elongated ($e\la0.2$). The counterparts of the INS candidates are also rather point-like, with mean ellipticity $\sim$\,0.08 -- the most elongated one being source \candtsf\ and the two objects possibly associated with \candtoe\ (see values in Table~\ref{tab_resmag}).

\subsubsection{Optical upper limits for \candzsf\label{sec_limmag}}
The only INS candidate of our sample that remains without an optical candidate is source \candzsf. The short ESO-VLT exposure obtained in 2007 revealed no counterparts brighter than $R\sim25$. The location of \candzsf\ towards the Carina star forming region explains the presence of many bright stars and of relatively intense diffuse emission in its optical field. The R filter, which includes the strong H$\alpha$ line, is thus particularly affected by background intensity variations, impacting the quality of the photometry.
In the light of this, we asked for additional deep SOAR exposures in the V band beginning of 2008. The V band excludes some of the strongest lines found in \ion{H}{ii} regions (Balmer emission as well as [\ion{N}{ii}] and [\ion{S}{ii}] emission lines) and is thus best suited to get a deep upper limit on the optical emission of \candzsf. However, the V filter is still somewhat contaminated by the nebular [\ion{O}{iii}] emission.

% ----- TABLE - FIT PARAMETERS OTHER CANDIDATES--------------------------------------------------------------
\begin{table*}
\begin{minipage}[t]{\textwidth}
\caption{X-ray spectral analysis of the INS candidates (excluding \candzsf)\label{tab_Xother}}
\centering
\renewcommand{\footnoterule}{}
\begin{tabular}{l l c c c c r c c c r c} 
\hline\hline
Candidate & OBSID & $t_{\rm exp}$\footnote{Exposure times are filtered for background flares.} & d.o.f & $kT$ & $f_{\rm X,14}$\footnote{Observed flux refers to range 0.15\,--\,3\,keV in units of $10^{-14}$\,erg\,s$^{-1}$\,cm$^{-2}$.} & $C$ & Goodness\footnote{The ``goodness-of-fit'' is derived from a number of 1000 Monte Carlo simulated spectra.} & $\Gamma$ & $f_{\rm X,14}$ & $C$ & Goodness \\
 & & (ks) &  & (eV) & (erg\,s$^{-1}$\,cm$^{-2}$) & & (\%) & & (erg\,s$^{-1}$\,cm$^{-2}$) & & (\%)\\
\hline
\candosf\  & 204710101 & 2.1  & 22 & $253^{+53}_{-40}$  & $3.6^{+0.8}_{-0.7}$    & 14 & 4.6  & 2.0(4) & $4.6^{+1.0}_{-0.9}$ & 17 & 9.9\\
\candtoe\  & 150870201 & 1.7  & 5  & $247^{+116}_{-60}$ & $2.6^{+0.9}_{-0.7}$    & 5  & 18.4 & 1.5(8) & $4.6^{+1.4}_{-1.3}$ & 5 & 19.8\\
\candtfoe\ & 307001301 & 7.3  & 24 & $200^{+29}_{-24}$  & $2.00^{+0.3}_{-0.29}$  & 22 & 26.8 & 2.4(4) & $2.8(4)$ & 21 & 22.7\\
\candtfie\ & 201750901 & 11.6 & 17 & $229^{+58}_{-44}$  & $1.7(3)$               & 24 & 76.7 & 2.1(4) & $2.5(5)$ & 21 & 60.0\\
\candtsf\  & 008830101 & 10.5 & 15 & $144^{+14}_{-12}$  & $2.00^{+0.22}_{-0.24}$ & 17 & 58.5 & $2.94^{+0.24}_{-0.23}$ & $2.59^{+0.3}_{-0.25}$ & 9 & 6.8\\
\candszf\  & 203020101 & 9.8  & 26 & $176^{+21}_{-18}$  & $2.15^{+0.26}_{-0.25}$ & 44 & 96.7 & $2.43^{+0.23}_{-0.22}$ & $3.1(3)$ & 35 & 79.4\\
\candseo\  & 303561001 & 7.8  & 11 & $186^{+34}_{-27}$  & $1.30^{+0.26}_{-0.24}$ & 6  & 6.7  & 2.6(4) & $1.7(3)$ & 6 & 7.0\\
\hline
\end{tabular}
\end{minipage}
\end{table*}
% -----------------------------------------------------------------------------------------------------------

To test the detection limit specifically in the region covered by the error circle of the X-ray source, small\footnote{$54''$\,$\times$\,$46''$, $128''$\,$\times$\,$100''$ and $70''$\,$\times$\,$70''$ respectively for the R, B and V images.} sections of the B, V and R images were analysed so as to minimize the contributions of nearby bright stars and of nebular emission. The frames were binned by a factor of 2 in order to increase the signal-to-noise ratio. Images of synthetic stars (created using the PSF model derived from the data) of progressively fainter magnitudes were added at the position of the X-ray source. The composite images were then subjected to automatic detection and to magnitude measurement using \textsf{DAOPHOT}. In this process, the synthetic star was no longer detected or was rejected (while trying to fit the PSF model) when the signal-to-noise ratio was worse than $S/N \sim 2.5$. We then defined the limiting magnitude as the magnitude of the faintest synthetic star still succesfully measured at this confidence level. We give in Table~\ref{tab_maglim} the magnitudes of the faintest simulated star ($m_{\rm syn}$) detected at a signal-to-noise ratio $\sigma_{\rm det}$, along with the measured magnitude ($m_{\rm det}$). The discrepancy between the simulated and detected values of the $R$ magnitude is due to a somewhat poorly defined PSF model in the wavelength range most affected by background intensity variations. Taking the limiting magnitudes as upper limits on the brightness of the optical counterpart of \candzsf, Table~\ref{tab_maglim} also lists the implied X-ray-to-optical flux ratios corrected for photoelectric absorption and interstellar extinction. In order to compute the unabsorbed fluxes we adopted the best blackbody fit parameters of the source (see Section~\ref{sec_X}) and derived $A_{\rm V} = 1.96$ using the \cite{pre95} relation between the X-ray absorption and optical extinction. For the other bands, we adopted the $A_\lambda/A_{\rm V}$ extinction relations of \cite{car89}.
% -----------------------------------------------------------------------------------------------------------

% ----- SECTION - X-RAY -------------------------------------------------------------------------------------
\section{X-ray analysis\label{sec_X}}

\candzsf\ is a very likely new thermally emitting INS with properties that, at first glance, seem similar to those of the \msev. A detailed X-ray and optical analysis of this object has recently been reported in \cite{pir08b}. We recall below the main results presented in this paper and refer to it for details on the reduction and analysis of the X-ray data of this and the other seven INS candidates.

\subsection{\candzsf}
\candzsf\ was detected altogether in 16 \xmm\ observations of the well studied binary system Eta Carinae and of the Wolf-Rayet star HD 93162 (WR\,25), as well as in one \chan\ observation of the Carina Nebula. Although the observing conditions were far from optimal -- in many observations \candzsf\ was located near the edge or in CCD gaps and often the effective exposure times were short -- we have analysed the available X-ray archival data on this source, which span more than six years. 

The event files were reprocessed using SAS 7.1.0 and CIAO 4.0.1\footnote{\texttt{http://cxc.harvard.edu/ciao4.0/index.html}}, applying standard procedure. Using XSPEC 12.4 \citep{arn96,arn04}, we tested different models (blackbody, power-law, bremsstrahlung, Raymond-Smith, \dots). All parameters were allowed to vary freely. Due to the low number of counts, we applied the C-statistic \citep{cas79} in order to derive the best fit parameters and their uncertainties. Whenever possible for a given observation, data from all EPIC cameras were analysed simultaneously to better constrain the spectral parameters.

We have shown that the X-ray emission of \candzsf\ is best described by a soft blackbody, with mean $kT\sim117\pm14$\,eV and $\nh\sim(3.5\pm1.1)$\,$\times$\,10$^{21}$\,cm$^{-2}$ and a stable 0.15\,--\,3\,keV observed flux of $f_{\rm X}\sim(1.03\pm0.06)$\,$\times$\,10$^{-13}$\,erg\,s$^{-1}$\,cm$^{-2}$ (errors are 3\,$\sigma$). There is no evidence for a hard non-thermal component. In this work we adopt the notation $F_{\lambda}$ to denote the unabsorbed fluxes while $f_{\lambda}$ is used for the observed flux. The $\nh$ is a factor of $\sim$\,10 higher than those typical of the \msev\ and is consistent with the one towards Eta Carinae, which is located at $\sim$\,2.3\,kpc \citep{smi06}. The distance to \candzsf\ is likely to be comparable. Together with the optical data, the high lower limit on the X-ray-to-optical flux ratio, $\log(F_{\rm X}/F_{\rm V})\ga3.1^{+0.3}_{-0.1}$, pratically excludes any other possibility than an INS. Timing analysis applying a Rayleigh $Z^2_n$ test shows no pulsations to a rather non-constraining 30\% upper limit (3\,$\sigma$), in the 0.073\,--\,100\,s period range. 

\subsection{Other INS candidates}
Each one of the other INS candidates was serendipitously observed only once by the \xmm\ detectors. The event files were reprocessed using SAS 8.0.0 applying standard procedure as for source \candzsf. For the spectral analysis, the low signal-to-noise ratio spectra were only fitted assuming an absorbed blackbody or power-law using XSPEC 12.4. 
Being roughly one order of magnitude fainter than \candzsf, the low number of counts prevents well constrained spectral fits; in particular, it does not permit a strong constraint on the value of the column density. Blackbody fits show rather high temperatures, typically $kT\ga200$\,eV, and column densities ranging from 0 to 4.2\,$\times$\,10$^{21}$\,cm$^{-2}$, at 68\% confidence level. Although apparently in disagreement with our selection criteria ($kT\le200$\,eV), the significant errors in HR explain the inclusion of these X-ray sources among the selected sources. We thus decided to hold $\nh$ fixed at the Galactic value (\citealt{dic90}, Table~\ref{tab_cand}) in order to obtain better constrained spectral fits. 
The results of the blackbody and power-law fits for these sources can be seen in Table~\ref{tab_Xother} (errors are 1\,$\sigma$).
The quality of each fit (``goodness'') corresponds to the fraction of simulations yielding a better fit statistic than the actual data, with high values implying bad fits.

Whereas the relatively large number of counts collected for \candzsf\ safely allows one to exclude a power-law shape for its X-ray spectrum, the much lower signal-to-noise spectra of the seven sources studied here are equally well described by either a hot blackbody or by a soft power-law energy distribution, with spectral indexes usually larger than 2.
% -----------------------------------------------------------------------------------------------------------

% ----- DISCUSSION ------------------------------------------------------------------------------------------
\section{Discussion\label{sec_disc}}

At energies below 2\,keV, the X-ray spectra of quasars and Seyfert I galaxies usually show an excess relative to the extrapolation of the high energy (2\,--\,10\,keV) power-law.
It is generally believed that this ``soft excess'' is produced by the scattering of thermal optical/UV photons from the accretion disk surrounding the central black hole to soft X-ray energies, by a population of ambient hot electrons \citep[e.g.][and references therein]{atl09}. 
Over a limited energy range -- in particular, that covered by the sensitivity of the \ros\ instruments, 0.1\,--\,2.4\,keV -- the energy distribution of the soft excess is roughly described by a power-law of spectral index $\ga$\,2 \citep[e.g.][]{bol96,gru98,pag04}. This is in agreement with the results obtained  in Section~\ref{sec_X}, for the INS candidates with optical counterparts (Table~\ref{tab_Xother}). Moreover, X-ray emitting quasars cluster in a region of the HR$_1$\,$\times$\,HR$_2$ diagram which is right above selection region III (contours in Fig.~\ref{fig_bestcand}), the location where it is expected to find sources with emission compatible with the hottest blackbodies that entered our selection. At faint fluxes, the 2XMMp sources show rather large HR errors, reflecting the fact that the low number of detected photons prevents a precise determination of the source spectral energy distribution. As it is evident from Fig.~\ref{fig_bestcand}, several of the INS candidates share hardness ratios common to both (galactic and extragalactic) populations.
For these reasons, our search procedure is sensitive to optically faint AGN, especially those showing soft X-ray excess.

Magnetic CVs exhibit a low temperature component detectable in soft X-rays or at UV wavelengths, depending whether the source is in a high or low accretion state \citep[e.g.][]{ram04a,ram04b}. This emission is due to the reprocessing of hard X-ray photons (of several tens of keV) produced in a shock of the accretion flow into the photosphere of the white dwarf. In principle, faint polars showing a particularly large soft X-ray excess and with uncatalogued optical counterparts can be present in our sample of INS candidates as well.

Active coronae of late-type stars manifest themselves as soft X-ray emission with luminosities generally below 10$^{31}$\,erg\,s$^{-1}$. This population dominates soft X-ray samples at low Galactic latitudes \citep{mot97a}.
It is well known, however, that the high latitude X-ray sky at flux level 10$^{-14}$\,erg\,s$^{-1}$\,cm$^{-2}$ is dominated by AGN \citep[e.g.][]{mac82,bar07,mat08}; the stellar content, although more important in soft X-ray-selected samples as ours, should be less than 10\%. In the optical, none of the INS candidates show colours typical of M, G, K stars (Fig.~\ref{fig_coldiag}). In any case, it is expected that late-type stars have $\log(f_{\rm{X}}/f_{\rm{R}})<-1$ \citep[e.g.][]{mac82,stocke91} which is clearly not the case for any of our INS candidates, all with $\log(f_{\rm{X}}/f_{\rm{R}})>0$.  Therefore, we do not expect any of the investigated X-ray sources to be identified with late-type stars. 

Based on the X-ray/optical associations discussed in Section~\ref{sec_Xopt}, the objects found inside the error circles of the INS candidates are very likely to constitute their true optical counterparts, all with a very low ($\la$\,5\%) probability of chance association, with the exception of \candtoe.
The strong UV excess exhibited by several of the sources (\candszf, \candtsf, and \candseo) is clearly consistent with those of AGN and CVs (Fig.~\ref{fig_coldiag}). It is worth noting, however, that optical colours alone would not be sufficient to distinguish these X-ray sources from the population of (X-ray and radio-quiet) blue objects located at high Galactic latitudes. In spite of the more average colours of the optical candidates for sources \candosf, \candtfoe\ and \candtfie, their high X-ray-to-optical flux ratios leave no doubt that they are likely of extragalactic origin. Still according to \cite{bar07}, the fraction of X-ray sources with $\log(f_{\rm{X}}/f_{\rm{R}})>1$, where obscured AGN are expected, although more important in hard X-ray-selected samples, is of just a few percent in soft ones. As mentioned in Section~\ref{sec_Xopt}, two out of seven of our candidates show $\log(f_{\rm{X}}/f_{\rm{R}})>1$, \candtoe\ and \candtsf. These are the sources with the most elongated optical candidates as well. In particular, the two optical objects found in the error circle of source \candtoe\ correspond to the most dubious of the X-ray/optical associations. They have been detected only in the R filter, preventing colour estimation, whereas the optical candidate of source \candtsf\ has been well detected in the U, B and R filters. \candtoe\ was detected in a very short exposure and therefore its HR errors are worse determined; however, the two possible optical counterparts define an already very high $\log(f_{\rm{X}}/f_{\rm{R}})\sim2$, characteristic of the most extreme classes of X-ray emitters as AM Her systems and BL Lac objects \citep{schwope99}, and it is worthy of further investigation.

Our best INS candidate is \candzsf, for which no optical counterpart is detected down to the limiting magnitude of our present data. On the basis of the analysis reported in \cite{pir08b}, the identification of \candzsf\ with an INS is very likely. This could be the first example of a presumably radio-quiet and X-ray dim INS, located at a significantly greater distance than the \msev. Its spatial location in the Galactic plane and the derived value of the column density suggest that it may be physically associated with the Carina Nebula, a giant \ion{H}{ii} region harbouring a large number of massive stars and with ongoing active star formation (see e.g. \citealt[and references therein]{smi07}). 

Several attempts aiming at identifying new thermally emitting INSs have been carried out in the years that followed the discovery of the seven sources, usually by cross-correlating \ros\ data with optical, radio and IR catalogues \citep[e.g.][]{rut03,agu06,chi05,tre07}. However, these searches are hampered by the large \ros\ positional errors at faint fluxes, especially in the populated regions of the Galactic plane. 
For these regions, many spurious (low significance) candidate optical/IR counterparts enter the X-ray error circle of a given \ros\ source. This makes the probability of erroneously assigning an identification significant, thus excluding the source as a potential INS candidate.
The final effect is that the Galactic plane is largely ``avoided'' by usual cross-correlation algorithms \citep[see e.g.][Fig.~1]{rut03}, and proposed INS candidates are usually located at high Galactic latitudes.

A long-term project which investigates INS candidates selected from the \ros\ Bright Source Catalogue \citep{vog99}, making use of follow-up investigations with the \swi\ satellite, is presently being conducted by R. E. Rutledge, D. B. Fox and collaborators \citep{rut08,let09}. 
Follow-up X-ray and optical observations of one candidate with no evident counterpart led to the discovery of Calvera, a likely INS (which exact nature is, however, still unclear) with a very large $F_{\rm X}/F_{\rm V}\ga8700$ \citep{rut08}. Not an exception, Calvera is located at high $b$, implying a remote distance of $d=8.4$\,kpc and a vertical velocity in excess of $\sim$\,5100\,km\,s$^{-1}$ to explain its current position well above the plane, considering standard cooling time and thermal emission similar to the \msev. Instead, it is more likely that Calvera is a nearby radio pulsar.

According to the detailed population synthesis calculations of \cite{pos08}, it is expected that, in general, new unidentified cooling INSs at faint soft X-ray fluxes are to be found in the Galactic plane, at small angular distances from their birth star forming regions -- in particular, of the rich OB associations located beyond the Gould Belt, as Carina, Vela, and Cygnus-Cepheus. At the flux limit applied in our search, the expected number of cooling neutron stars having the same properties as those exhibited by \candzsf\ -- i.e. a slightly hotter source located at a greater distance relative to the \msev\ -- is of only $\sim$\,50\,--\,80 in the whole sky\footnote{The range shown here for the expected number of sources is derived from \cite{pos08} adopting the dotted and hashed curves of their Fig.~5 as lower and upper limits, respectively. See \citealt{pos08} for details on the assumed parameters. EPIC pn counts were converted to \ros\ PSPC counts using WebPIMMS (\texttt{http://heasarc.gsfc.nasa.gov/Tools/w3pimms.html}).}, which translates into less than one source to be present in the 2XMMp catalogue. 
The fact that we found one source exhibiting these unique characteristics only corroborates \cite{pos08} conclusions that a search for new cooling INSs should not be blind, i.e. one has to look for preferentially in the most promising regions of the sky -- i.e. near more remote OB associations.

Population synthesis constitutes a fundamental tool in order to compare observational results from e.g. surveys with theoretical expectations for a given population of astrophysical objects. If a large number of objects are known, as for the case of radio pulsars, it allows testing many of the population physical (and theoretically unknown) properties. 
Recent examples are the work of \cite{fau06}, conducted for the population of radio pulsars, and the study carried out by \cite{sto07} with the population of disk millisecond radio and $\gamma$-ray pulsars.

In the particular case of thermally emitting INSs, population syntheses conducted in the early 90's predicted that a large number of old neutron stars accreting from the ISM would be detected by \ros\ \citep{tre91,bla93,mad94}. It soon became clear that the unrealistic overprediction was due to several simplfying assumptions used in the model.
In fact, several mechanisms act in order to inhibit accretion such as the presence of the magnetic field of the neutron star \citep{tor03,ikh07} and that of an infalling material which is weakly magnetized \citep{per03} or heated by the emergent X-rays \citep[e.g.][]{bla95}. In any case, it is believed that accretion should proceed at a rate well below the Bondi-Hoyle one, thus affecting the actual number of sources that can be observed by current X-ray missions.
The population of cooling nearby INSs from the Gould Belt has been extensively modeled by \cite{pop00b,pop03,pop05}. In \cite{pop06}, the expected number of sources at a given flux was used to constrain cooling curves of INSs. \cite{pos08} improved many of the features of their population synthesis model, mainly by providing a new description of the ISM distribution and that of the massive progenitors. Interestingly, they used the spatial distribution of the synthetic population in order to define search strategies to identify new cooling INSs. 

The expected number of such sources to be detected by current X-ray missions is strongly model-dependent. It involves from the parametrization of properties inherited at birth (kick velocity and mass distribution of the compact remnant) to that of the cooling history and emission mechanism of the neutron star. Together with realistic descriptions of the Galactic gravitational potential, the ISM distribution and the detector characteristics, these ingredients determine the neutron star thermal and dynamic evolution and detectability. Similarily, from the observational point of view, selection biases and the area effectively covered by the survey (considering the constraints applied while correlating X-ray sources with other wavelengths) have to be taken into account in order to derive correct observational upper limits on the number of sources at a given flux. 
Overall, even the simplest models end up having a level of complexity which is made evident from the number of assumed hypotheses -- many of them involving significant theoretical and observational uncertainties. We are currently developping a model for this Galactic population which aims at making the least number of uncertain assumptions -- even at the expense of having a less sophisticated description of the real population -- while, based on Monte Carlo simulations and leaning on what it is observed, in particular, with the \xmm, would be statistically robust in order to constrain important properties of this population of unique Galactic remnants especially beyond the Solar vicinity. 
% -----------------------------------------------------------------------------------------------------------

% ----- SUMMARY & CONCLUSIONS -------------------------------------------------------------------------------
\section{Summary and conclusions\label{sec_conc}}

Our search for new thermally emitting INSs in the 2XMMp catalogue has revealed a handful of interesting and previously unknown soft X-ray sources. Deep optical imaging revealed likely optical counterparts for six of them which, based on blue optical colours and X-ray-to-optical flux ratios around 1\,--\,10, identify them with, most likely, AGN and CVs. Source \candtoe\ has no evident optical candidate and should be further investigated. 

Source \candzsf\ is a newly discovered thermally emitting INS, possibly radio-quiet and similar to the \msev. As it is expected for cooling neutron stars, its X-ray emission is described by an intrinsically soft and thermal energy distribution, stable on long time scales and with no evidences for magnetospheric activity. The present lower limit on the X-ray-to-optical flux ratio already excludes standard classes of X-ray emitters. Its likely location in the Carina Nebula is in agreement with the most up-to-date expectations of population synthesis models.

This work confirms the use of the \xmm\ catalogue of sources as an efficient tool in order to identify new thermally emitting INSs and other particular classes of soft X-ray sources. A similar search using the \chan\ data with its excellent astrometric accuracy would be instrumental to select sources especially located in the Galactic plane, where it is more likely to find new INS candidates.
% -----------------------------------------------------------------------------------------------------------

% ----- ACKNOWLEDGEMENTS ------------------------------------------------------------------------------------
\begin{acknowledgements}
We thank N. Grosso for discussions about \chan\ data and source \candzsf. The work of A. M. P. is supported by FAPESP (grant 04/04950-4), CAPES (grant BEX7812/05-7), Brazil, and the Observatory of Strasbourg (CNRS), France.
\end{acknowledgements}
% -----------------------------------------------------------------------------------------------------------

% ----- BIBLIO ----------------------------------------------------------------------------------------------
\bibliographystyle{aa}
\bibliography{ins}
% -----------------------------------------------------------------------------------------------------------

\end{document}